
\input harvmac
\def\etal{\it et al. \rm}
\def\bra#1{\left\langle #1\right|}
\def\ket#1{\left| #1\right\rangle}

\def\vev#1{\left\langle #1\right\rangle}
\def\hf{{\textstyle{1\over2}}}
\def\thf{{\textstyle{3\over2}}}
\def\sst{\scriptscriptstyle}
\def\sst{\scriptstyle}

\def\sqr#1#2{{\vcenter{\vbox{\hrule height.#2pt
        \hbox{\vrule width.#2pt height#1pt \kern#1pt
           \vrule width.#2pt}
        \hrule height.#2pt}}}}

\def\refmark#1{${}^{\refs{#1}}$\ }
\def\footsym{*}\def\footsymbol{}\ftno=-2
\def\foot{\ifnum\ftno<\pageno\xdef\footsymbol{}\advance\ftno by1\relax
\ifnum\ftno=\pageno\if*\footsym\def\footsym{$^\dagger$}\else\def\footsym{*}\fi
\else\def\footsym{*}\fi\global\ftno=\pageno\fi
\xdef\footsymbol{\footsym\footsymbol}\footnote{\footsymbol}}
\lref\RajWein{R. Rajaraman and E. Weinberg,
\sl Phys.~Rev.~\bf D11\rm, 2950 (1975).}
\lref\ItZub{See for example C. Itzykson  and J. B. Zuber,
\sl Quantum field theory\rm, NY, McGraw-Hill, 1980, Sec.~5-1-2.}
\lref\Holzwarth{G. Holzwarth, G. Pari and B. K. Jennings,
\sl Nucl.~Phys.~\bf A515\rm, 665 (1990);
G. Holzwarth, \sl Phys.~Lett.~\bf B241\rm, 165 (1990) and pp.~279-302,
in G. Holzwarth, ed., {\sl Baryons as Skyrme Solitons}, $ibid.$}
\lref\VerMore{H. Verschelde, \sl Phys.~Lett.~\bf B249\rm, 175 (1990);
\sl Nucl.~Phys.~\bf A523\rm, 563 (1991).}
\lref\DiakNew{D. Diakonov, \sl Acta Phys.~Pol.~\bf B25\rm, 17 (1994);
D. Diakonov and V. Petrov, St.~Petersburg preprint LNPI-1394,
1988 (unpublished).}
\lref\OhtaMore{K. Ohta, \sl Nucl.~Phys.~\bf A511\rm, 620 (1990);
\sl Phys.~Lett.~\bf B242\rm, 334 (1990).}
\lref\Ringwald{A. Ringwald, \sl Nucl. Phys. \bf B330\rm, 1 (1990).}
\lref\Ramond{See for example P. Ramond, \sl Field theory: a modern
primer\rm, NY, Benjamin/Cummings, 1981, pp.~74-80.}
\lref\Soldate{M. Soldate, \sl Int.~Journ.~Mod.~Phys.~\bf E1\rm, 301
(1992).}
\lref\Schroers{B. J. Schroers, \it Zeit.~Phys.~\bf C61\rm, 479 (1994), Sec.~3.}
\lref\RebSlan{C. Rebbi and R. Slansky, \sl Rev.~Mod.~Phys.~\bf42\rm,
68 (1970).}
\lref\ANW{G. Adkins, C. Nappi and E. Witten, {\sl Nucl. Phys. }
{\bf B228} (1983) 552; G. Adkins and C. Nappi, {\sl Nucl. Phys. }
{\bf B233} (1984) 109.}
\lref\AmAlg{R. D. Amado, R. Bijker and M. Oka,
\sl Phys.~Rev.~Lett.~\bf58\rm, 654 (1987); M. Oka \it et al.\rm,
\sl Phys.~Rev.~\bf C36\rm, 1987 (1727); M. Oka, \sl Phys.~Lett.~\bf
205B\rm, 1 (1988); R. D. Amado, M. Oka and M. P. Mattis,
\sl Phys.~Rev.~\bf D40\rm, 3622 (1989).}
\lref\IJrule{M. Mattis and M. Mukerjee, \sl Phys.~Rev.~Lett.~\bf61\rm,  1344
 (1988); M. Mattis and E. Braaten, \sl Phys.~Rev.~ \bf D39\rm, 2737  (1989);
M. Mattis, \sl Phys.~Rev.~ \bf D39\rm, 994 (1989);
\sl Phys.~Rev.~Lett.~\bf 63\rm, 1455 (1989).}
\lref\RW{ R. Rajaraman and E. Weinberg, {\sl Phys. Rev.} {\bf D11}
(1975) 2950.}
\lref\Skyrme{ T. H. R. Skyrme, {\sl Proc. Roy. Soc.} {\bf A260} (1961) 127;
\sl Nucl.~Phys.~\bf31 \rm (1962) 556.}
\lref\DiakPet{ D. I. Dyakonov, V. Yu. Petrov, and P. B. Pobylitsa,
{\sl Phys. Lett.} {\bf B205} (1988) 372.}
\lref\HEHW{A. Hayashi, G. Eckart, G. Holzwarth and H. Walliser,
{\sl Phys. Lett.} {\bf B147} (1984) 5; B. Schwesinger, H. Weigel,
G. Holzwarth and A. Hayashi, \sl Phys.~Rep.~\bf173\rm, 173 (1989).}
\lref\MK{M. Mattis and M. Karliner {\sl Phys. Rev.} {\bf D31} (1985)
2833; M. Mattis and M. Peskin, {\sl Phys. Rev.} {\bf D32} (1985) 58;
M. Karliner and M. Mattis, \sl Phys.~Rev.~\bf D34\rm, 1991 (1986);
M. Mattis, \sl Phys.~Rev.~Lett.~\bf 56\rm, 1103 (1986).
As argued convincingly in Ref.~\KKKO\ (and contrary to the claims
of Ref.~\DiakPet\ among others), it is likely that the potential
scattering phase-shift analysis of Refs.~\HEHW-\MK\ carries information,
not only about
$\langle\delta\pi^a(x)\,\delta\pi^b(y)\rangle$ (Eq.~\secondcont),
but also about the ``Compton diagram'' contribution \schema, albeit
only in the strict large-$N_c$ limit where the $\Delta$ is degenerate
with the nucleon so that the resonance contribution to the $S$ matrix
is lost.}
\lref\Goldstone{ J. Goldstone and R. Jackiw, {\sl Phys. Rev.} {\bf D11}
(1975) 1486. See also C. Coriano \it et al\rm., \it ibid\rm. \bf 45\rm,
2542 (1992).}
\lref\CG{ C. Callan and D. Gross, {\sl Nucl. Phys.} {\bf B 93} (1975) 29.}
\lref\GSII{ J. Gervais and B. Sakita, {\sl Phys. Rev.} {\bf D30}
(1984) 1795.}
\lref\DashMan{R. Dashen and A. Manohar,
{\sl Phys. Lett.} {\bf B315} (1993) 425.}
\lref\Hajduk{C. Hajduk and B. Schwesinger, \sl Nucl.~Phys.~\bf A453 \rm
(1986) 620; V. G. Makhankov, Yu.~P. Rybakov and V. I. Sanyuk,
\it Sov.~Phys.~Usp.~\bf 35 \rm (2), 55, Sec.~3.6.}
\lref\Uehara{M. Uehara, \sl Prog.~Theor.~Phys.~\bf 75 \rm (1986) 212;
\sl ibid., \bf 78 \rm (1987) 984.}
\lref\Adami{C. Adami and I. Zahed,  {\sl Phys. Lett.} {\bf B213} (1988) 373.}
\lref\Verschelde{ H. Verschelde, {\sl Phys. Lett.} {\bf B209}
(1988) 34; {\sl Phys. Lett.} {\bf B215} (1988) 444;
H. Verschelde and H. Verbeke, \sl Nucl.~Phys.~\bf A495 \rm (1989) 523.}
\lref\HHS{ G. Holzwarth, A. Hayashi and B. Schwesinger, {\sl Phys. Lett.}
{\bf B191} (1987) 27.}
\lref\Otofuji{ T. Otofuji, S. Saito and M. Yasuno,
 {\sl Prog. Theor. Phys.} {\bf 73} (1985) 520. DO WHAT SCHNITZER DOES}
\lref\Saito{S. Saito, {\sl Prog. Theor. Phys.} {\bf 78} (1987) 746;
{\sl Nucl. Phys.} {\bf A463} (1987) 169c.}
\lref\Ohta{K. Ohta, {\sl Phys. Lett.} {\bf B234} (1990) 229.}
\lref\Oka{M. Oka, H. Liu and R. D. Amado, {\sl Phys.~Rev.} {\bf C39}
 (1989) 2317.}
\lref\GJS{ J. Gervais, A. Jevicki and B. Sakita, {\sl Phys. Rev.}
{\bf D12} (1975) 1038.}
\lref\GS{ J. Gervais and B. Sakita, {\sl Phys. Rev.} {\bf D11} (1975) 2943.}
\lref\deVone{ H. de Vega, J. Gervais and B. Sakita {\sl Nucl. Phys.}
{\bf B142} (1978) 125.}
\lref\deVtwo{ H. de Vega, J. Gervais and B. Sakita {\sl Phys. Rev.}
{\bf D19} (1979) 601.}
\lref\Tomboulis{ E. Tomboulis, {\sl Phys. Rev.} {\bf D12} (1975) 1678.}
\lref\GJ{ J. Gervais and  A. Jevicki, {\sl Nucl. Phys.} {\bf B110}
(1976) 93.}
\lref\schoolbook{ L. S. Schulman, {\sl Techniques and Applications of Path
Integration}, (Wiley-Interscience 1981). }
\lref\Berezin{ F. A. Berezin, {\sl Theor. Math. Phys.} {\bf 6} (1971) 194. }
\lref\Coleman{ S. Coleman, {\sl Phys. Rev.} {\bf D11} (1975) 2088. }
\lref\ZB{I. Zahed and G. E. Brown, \sl Phys.~Rep.~\bf142 \rm (1986) 3;
K. F. Liu, ed.,  {\sl Chiral Solitons},  Singapore,
World Scientific Press, 1987; G. Holzwarth., ed., {\sl Baryons as
Skyrme Solitons}, Singapore, World Scientific Press, 1992;
V. G. Makhankov, Y. P. Rybakov and V. I. Sanyuk, \sl The Skyrme Model:
fundamentals, methods, applications\rm, NY, Springer-Verlag, 1993.}
\lref\KKKO{K. Kawarabayashi and K. Ohta, \sl Phys.~Lett.~\bf B216\rm,
205 (1989). The results of this illuminating paper are corrected and
extended in  Y. G. Liang \etal, \sl Phys.~Lett.~\bf B243\rm, 133 (1990).}
\lref\DHM{N. Dorey, J. Hughes and M. Mattis, \sl Phys.~Rev.~\bf D49\rm, 3598
(1994).}
\lref\Rajetal{R. Rajaraman, H. M. Sommerman, J. Wambach and
H. W. Wyld, \sl Phys.~Rev.~\bf D33\rm, 287 (1986).}
\lref\Bander{M. Bander and F. Hayot, \sl Phys.~Rev.~\bf D30\rm, 1837 (1984).}
\lref\Liuetal{B. A. Li, K. F. Liu and M. M. Zhang,
\sl Phys.~Rev.~\bf D35\rm, 1693 (1987); K. F. Liu, J. S. Zhang
and G. R. E. Black, \sl Phys.~Rev.~\bf D30\rm, 2015 (1984).}
\lref\Braaten{E. Braaten and J. Ralston, \sl Phys.~Rev.~\bf D31 \rm (1985)
598.}
\lref\HSU{A. Hayashi, S. Saito and M. Uehara,
{\sl Phys. Lett.} {\bf B246} (1990) 15;
 \sl Phys.~Rev.~\bf D43\rm, 1520 (1991); \it ibid\rm., \bf D46\rm, 4856
(1992); \sl Prog.~Theor.~Phys.~Supp.~\bf109 \rm (1992) 45;
H. Riggs and H. Schnitzer, {\sl Phys. Lett.} {\bf B305} (1993) 252.}
\lref\LY{H. Levine and L. Yaffe, \sl Phys. Rev. \bf D19\rm, 1225 (1979).}
\lref\Schulman{L. S. Schulman, \sl Phys. Rev. \bf 176\rm, 1558 (1968).}
\lref\GasLSZ{S. Gasiorowicz, \sl Elementary particle physics\rm, New York,
Wiley, 1966.}
\lref\Gas{H. T. Williams,  \sl Phys.~Rev.~\bf C31\rm, 2297 (1985).}
\lref\Peccei{R. Peccei, \sl Phys.~Rev.~\bf 176\rm, 1812 (1968).}
\lref\otherspin{R. E. Behrends and C. Fronsdal,
\sl Phys.~Rev.~\bf 106\rm, 345 (1957).}
\lref\FadKor{L. D. Faddeev and V. E. Korepin, \sl Phys.~Rep.~\bf42 \rm
(1978) 1.}
\lref\ArnMat{N. Dorey, J. Hughes and M. Mattis,
hep-ph@xxx.lanl.gov/9406406, Physical Review Letters (in press);
P. Arnold and M. Mattis, \sl Phys.~Rev.~Lett.~\bf65\rm,
  831 (1990).}
\lref\ManoharRG{A. Manohar, UCSD/PTH 94-14, hep-ph@xxx.lanl.gov/9407211.}
\lref\Witten{ E. Witten, \sl Nucl.~Phys.~\bf B160 \rm
(1979) 57.}
\font\authorfont=cmcsc10 \ifx\answ\bigans\else scaled\magstep1\fi
{\divide\baselineskip by 3
\multiply\baselineskip by 2
\def\prenomat{\matrix{&\scriptstyle\rm hep-ph/9404274\cr
&\scriptstyle\rm Phys.~Rev.~\bf D50\rm, 5816 (1994)}}
\Title{$\prenomat$}{\centerline{Skyrmion Quantization and the
Decay of the $\Delta$}}
\centerline{\authorfont Nicholas Dorey}
\bigskip
\centerline{\sl Physics Department, University College of Swansea}
\centerline{\sl Swansea SA2$\,$8PP UK $\quad$ \rm pydorey@pygmy.swan.ac.uk}
\bigskip
\centerline{\authorfont James Hughes}
\bigskip
\centerline{\sl Physics Department, Michigan State University}
\centerline{\sl East Lansing, MI 48823 USA$\quad$ \rm hughes@msupa.pa.msu.edu}
\bigskip
\centerline{\authorfont Michael P. Mattis}
\bigskip
\centerline{\sl Theoretical Division T-8, Los Alamos National Laboratory}
\centerline{\sl Los Alamos, NM 87545 USA$\quad$ \rm mattis@skyrmion.lanl.gov}
\vskip .3in
\noindent
We present the complete solution to the so-called ``Yukawa problem'' of
the Skyrme model.  This refers to the perceived difficulty of
reproducing---purely from soliton physics---the usual pseudovector
pion-nucleon coupling, echoed by pion coupling to the higher
spin/isospin baryons $\textstyle{\big(I=J={3\over2},{5\over2},\cdots,
{N_c\over2}\big)}$ in a manner fixed by large-$N_c$ group theory.
The solution involves surprisingly elegant interplay between the classical
and quantum properties of a new configuration: the \it rotationally
improved skyrmion\rm. This is the near-hedgehog obtained by minimizing
the usual skyrmion mass functional augmented by an all-important
(iso)rotational kinetic term. The numerics are pleasing: a $\Delta$ decay
width within a few $\,$MeV of its measured value, and furthermore, the
higher-spin baryons $\textstyle{\big(I=J\ge{5\over2}\big)}$ with widths
so large ($\Gamma>800\,$MeV) that these undesirable large-$N_c$
artifacts effectively drop out of the spectrum, and pose no
phenomenological problem. Beyond these specific results, we ground the
Skyrme model in the Feynman Path Integral, and set up a transparent
collective coordinate formalism that makes maximal use of the $1/N_c$
expansion. This approach elucidates the connection between skyrmions on
the one hand, and Feynman diagrams in an effective field theory on the
other.\looseness=-1\relax
\vskip .1in
\Date{\bf April 1994} 
\vfil\break
}

\newsec{Overview}

Since its reinvention a decade ago by Adkins, Nappi and
Witten,\refmark\ANW and despite
many phenomenological successes,\refmark\ZB the Skyrme
model\refmark\Skyrme
 has suffered from too many competing, often conflicting, formalisms.
Particular confusion surrounds those problems that involve
interactions between skyrmions (read: baryons) and the elementary quanta of
the theory (read: mesons). In fact, the most basic such question one
can pose---how does the Skyrme model generate the correct pion-baryon
3-point coupling?---has not been satisfactorily resolved.

In this paper, we present the complete solution to this so-called
``Yukawa problem,'' as it has come to be known in the literature.
Both the problem and the solution are detailed in this expanded introductory
Section, which also sets forth some general principles that we think
are important, and underappreciated.

This is the second in a series of papers intended to clarify the
nature of the meson-skyrmion interactions, by grounding the Skyrme model
in the Feynman Path Integral (FPI). The small parameter of the Skyrme
model is $1/N_c$ ($N_c$ being the number of colors of the underlying
gauge theory) which is tied to the loop expansion natural to the FPI
by appearing in the combination $\hbar/N_c.$ In Ref.~\DHM\ we focused
on a ``toy'' version of the Skyrme model, in which space-time is
1+1 dimensional, and the internal flavor symmetry is $U(1)$ rather than
chiral $SU(2)\times SU(2).$ The process we analyzed, while
simple, proved  illuminating: the decay of a soliton in its
$n$th excited state, to its next-lower state, by emission of a single charged
meson. Here we will focus not only on the analogous physical decay
$\Delta\rightarrow N\pi$, but also on virtual processes such as
$N\rightarrow N\pi$ and $\Delta\rightarrow \Delta\pi$ that are
building blocks for more complicated diagrams, and likewise for
all the higher spin/isospin baryons ($I=J=\textstyle{5\over2},$
$\textstyle{7\over2},$ etc.) that emerge as
rotational excitations of the hedgehog skyrmion.  While numerics
are not our primary goal at the present, it is  pleasing that the
width of the $\Delta$ in the Skyrme model works out to 114 MeV versus
120$\pm$5 MeV experimentally (a result not original to us, but rather
confirming a large-$N_c$ ansatz in Ref.~\ANW), while the higher-spin
baryons are so broad ($>800$ MeV) that they would not normally
be classified as ``particles''---here again, by their absence from
the spectrum, in agreement with Nature.\foot{See Sec.~6
below for width calculations.}

By virtue of the delicate interplay between its classical and quantum
 properties, the Skyrme model will be seen to be richer and more elegant
than the $U(1)$ toy model. Yet the three main points of Ref.~\DHM, which
one might characterize as a \it caveat\rm, a \it prescription\rm, and
a \it moral\rm, hold here as well, and provide useful guideposts for
the development below.  We review them accordingly:

\def\vphi{\vec\phi}
\def\fpi{f_\pi}
\def\vpi{\vec\pi}
\def\cl{{\rm cl}}
\def\e{\epsilon}
\def\eps{\e}
\def\mpi{m_\pi}
\def\tot{{\rm tot}}
\it A caveat\rm. Whether in the $U(1)$ model where the elementary boson
is the real scalar doublet $\vphi,$ or in the Skyrme model where it is the
pion field $\vpi,$ or in other soliton theories,
it is customary to split up the total field into
classical and fluctuating parts, $\vphi=\vphi_\cl+\delta\vphi.$
Here $\vphi_\cl$ is the classical soliton, whose zero modes are the
``baryon degrees of freedom,'' while the fluctuating field $\delta\vphi,$
properly orthogonalized to these zero modes, is often said to represent
the ``meson degrees of freedom.'' Unfortunately, this commonly held
distinction between ``baryon'' and ``meson'' degrees of freedom is
\it false\rm, and leads to incorrect
results.\foot{\divide\baselineskip by 3\multiply\baselineskip by 2
In the particular example of Yukawa couplings in the Skyrme model,
we find ourselves disagreeing with
Refs.~\refs{\Verschelde\Saito\HHS\Uehara\Adami-\Ohta},
among many others, who
identify as the physical pion only the  fluctuating  field $\delta\vpi$.
Since $\delta\vpi\sim N_c^0$
while $\vpi_\tot\propto\fpi\sim N_c^{1/2},$ it is no surprise, and has been
noted by several of these authors,  that the width
of the $\Delta$ as may be calculated in these formalisms is down by (at
least) one power of $N_c$ from the unambiguous, model-independent leading-order
answer derived from  Eq.~(1.3)
 below. Phrased the usual way, since the first variation $\delta\vpi\cdot
\big(\delta S/\delta\vpi\big)$ off the \it static \rm skyrmion vanishes
by the defining equation, the first variation off the \it rotating
\rm skyrmion ($i.e.,$ the nucleon or $\Delta$)
is suppressed in large $N_c,$ since the skyrmion rotates slowly
($\omega_{\rm rot}^{}\sim N_c^{-1}$). The more complicated case of
pion-skyrmion scattering will be discussed in Sec.~7.}
  While it may be convenient
mathematically to split up the field in this way at an intermediate
stage in the calculation (as we ourselves do in Secs.~2-3 below), physically
meaningful Green's functions $\cal G$
must be formed from the reconstituted \it total
\rm field, $\vphi$ or $\vpi.$ The presence, or absence, of
asymptotic states containing $n$ physical
mesons with 4-momenta $q_1,\cdots,q_n$ can then be gleaned from
the analytic properties of $\cal G$, as per the LSZ amputation
 procedure.\foot{A thorough review of LSZ is chapter 7 of Ref.~\GasLSZ.
The observation that the classical part of the field cannot be ignored,
and on the contrary contributes at leading order to poles in the ``meson''
channels, dates back to the early literature on soliton
quantization.\refmark{\Goldstone,\CG}}
 Specifically, one looks for simultaneous  poles on the meson mass shell,
${\cal G}\sim\prod_{i=1}^n\big(q_i^2-\mpi^2+i\eps)^{-1}+\cdots,$
and identifies the residue with the  $S$ matrix element.
 As for the baryons/skyrmions, the  LSZ procedure
is even simpler, since in large $N_c$ baryons are very heavy
(masses $\sim N_c$) and can for present purposes
be treated in a nonrelativistic, first-quantized manner. To reiterate,
we concentrate on these analytic properties, not because we are
infatuated with formalism, but because we want to avoid wrong answers.

\def\I{{\cal I}}
\def\M{{\rm M}_s}
\it A prescription\rm. As stated above,  $N_c$ enters into the problem in the
combination $N_c/\hbar$.
An economical formalism should exploit this fact, and map a leading-order
calculation in $1/N_c$ onto a \it zeroth \rm order expression,
$i.e.$ a saddle-point, in the
semiclassical expansion (and not some ``higher-order effect''
in a naive perturbative expansion as
often appears in the skyrmion literature when the author
elects to split up the total pion
 field).  Framed in these terms, the problem is that,
for purposes of LSZ, $\vphi_\cl$ is the \it wrong approximate saddle-point\rm.
By definition, $\vphi_\cl$ solves
the static Euler-Lagrange equation $0=\delta \M/\delta\vphi$ where
$\M[\vphi,\partial_x\vphi\,]$ is the soliton mass functional
(the integrated Hamiltonian). The prescription we proved in Ref.~\DHM\ is to
solve, instead, the static equation
$0 =\delta\big(\M+P^2/2\I\,\big)/\delta\vphi$
where $P$ is the momentum conjugate to the soliton's $U(1)$
collective coordinate $\theta,$ and $\I[\vphi\,]$ is the soliton's
moment of inertia. Likewise, in the Skyrme model, we will show that the
right semiclassical starting point is the solution to
\eqn\riskySUtwo{0\ =\ {\delta\over\delta\vpi}\left(\M+\hf J^m\I^{-1}_{mn}J^n\,
\right)}
where $\I$ is now a tensor. We  call such solutions
\it rotationally improved skyrmions\rm, and they are no longer
precisely hedgehogs (a key point).\foot{A similarly distorted hedgehog
has recently been obtained by Schroers using different
methods.\refmark\Schroers A technical aside:
It is widely believed that when the skyrmion is not precisely a hedgehog,
one must in principle
introduce extra collective coordinates for $iso$rotations in
addition to spatial rotations, since these are no longer equivalent,
and concomitantly, an additional
$iso$rotational
 kinetic energy term beyond the one displayed in Eq.~\riskySUtwo.
But for the rotationally improved skyrmion,
 our FPI formalism clarifies that \it this is not the case\rm;
Eq.~\riskySUtwo\ suffices (see Appendix C for a discussion).
In addition, the FPI approach completely obviates an ongoing dialectic
about the relative merits of this or that
``gauge'' (meaning how one chooses to orthogonalize the fluctuating
modes from the skyrmion's zero modes). To emphasize this point, Secs.~2-3
below are framed in the most general (linear) gauge, and the
gauge invariance of our physical results is manifest.}
Ostensibly, the added rotational kinetic term is a small perturbation,
since $\I^{-1}\sim N_c^{-1}.$ Nevertheless its effect on the analytic structure
of the Green's functions is critical, and cannot be neglected.
In the $U(1)$ toy model it contributes a negative mass-squared, so that
rather than falling off as $\exp-m^{}_\phi r$ like $\vphi_\cl,$ the
rotationally improved soliton
$\sim\,\exp-{(m^2_\phi-P^2/\I^2)}_{}^{1/2} r$. In momentum space this turns
$\big(|{\bf q}|^2+m^2_\phi\big)^{-1}$ into
$\big(|{\bf q}|^2+m^2_\phi-P^2/\I^2\big)^{-1}$. As
 $P/\I$ can be equated to the meson energy, ${\cal G}(q)$
now correctly has a pole on the mass shell, and excited $U(1)$
solitons can legally
decay by meson emission. In the Skyrme model, the effect of the rotational
perturbation is more interesting. Thanks to its deviation away
from the hedgehog ansatz, the rotationally improved skyrmion
falls off as a superposition of two distinct
exponentials, so that in momentum space
\eqn\mombehavior{{\cal G}(q)\ \sim\
{{\cal N}_1\over|{\bf q}|^2+m^2_\pi- J^2/\I^2}\ +\
{{\cal N}_2\over|{\bf q}|^2+m^2_\pi}\ +\ \hbox{(non-pole terms)}\ .}
This makes perfect sense:
the first pole  correctly describes skyrmion decay processes such
as $\Delta\rightarrow N\pi$ while the second describes
 $N\rightarrow N\pi$, etc!\foot{Very roughly speaking,
the arithmetic works as follows. The pion energy $\omega_\pi^{}$ must equal
the difference of the initial and final skyrmion energies,
namely $J_i(J_i+1)/2\I\,-\,J_f(J_f+1)/2\I.$ When $J_f=J_i-1$ this difference is
$J_i/\I$ whereas when $J_f=J_i$ it is zero,
consistent with the two pole locations in Eq.~\mombehavior,
respectively. The actual
analysis of Sec.~5 is not quite so simple:  operator ordering
ambiguities must be resolved.} In either model, this pole-shift
phenomenon is a variation on the old exercise of expanding a field
theory about the wrong mass, and treating the mass-shift in perturbation
theory: an increasing number of 2-point insertions must be summed
geometrically (as accomplished implicitly in Eq.~\riskySUtwo) to move
the pole in the propagator to the physical mass-shell.

\it A moral\rm. The moral of Ref.~\DHM\ is equally valid for the Skyrme
model, namely, the order-by-order equivalence of the soliton theory
to an effective relativistic quantum field theory with
explicit baryon fields.\foot{Significant progress
 in the reverse direction, from field theory to skyrmions, can
be found in Refs.~\ArnMat-\ManoharRG, in which the Skyrme model (or
variants thereof) is conjectured to emerge as an ultraviolet
renormalization group fixed point of
a class of effective meson-baryon Lagrangian field theories.}
 The eventual goal
of mapping out this effective theory in full is well beyond the scope of
the present paper. But by focusing on
skyrmion decay by one-meson emission, we shed light on the effective
3-point meson-baryon vertex which can, in turn, be assembled into
more complicated Feynman diagrams ($e.g.,$ $\pi N$ scattering, or
pion-exchange contributions to the $NN$ system). In the $U(1)$ model, the
baryon/soliton states $\ket{p}$ are labeled by an integer charge.
The baryon wavefunctions are $\psi_p^{}(\theta)=\vev{\theta|p}
=e^{ip\theta},$ and their effective Yukawa couplings to the charged scalars
$\phi^\pm\equiv\phi_1\pm i\phi_2$ can be expressed as
$g\phi^+(x)\int d\theta\,e^{i\theta}\ket{\theta}\bra{\theta}+
\hbox{h.c.}$, or even more compactly as
$g\phi^+(x)e^{i\hat\theta}+\hbox{h.c.}$, where $\hat\theta\ket{\theta}=
\theta\ket{\theta}.$
The presence of $e^{i\theta}$ properly insures $\Delta p=1,$
as the meson carries away one unit of charge.
In 3+1 dimensions, with $SU(2)\times SU(2)$ symmetry,
the form of the analogous
pion-baryon effective coupling  is \it completely determined
a priori \rm by the
twin requirements of the chiral and large-$N_c$ limits, and
reads\foot{This coupling, which we review in Appendix D,
 was first written down in Sec.~5 of Ref.~\ANW,
with no explicit input from or connection to skyrmion physics.}
\def\D{D^{\scriptscriptstyle(1)}}
\def\gpnn{g_{\pi\scriptscriptstyle NN}^{}}
\def\gpnnsqd{g_{\pi\scriptscriptstyle NN}^2}
\def\gpnd{g_{\pi\scriptscriptstyle N\Delta}^{}}
\eqn\jointreq{-{3\gpnn\over2 M_N}\partial_i\pi^a\int_{SU(2)}dA\,
\D_{ai}(A)\ket{A}\bra{A}\ +\ \cdots\ ,\quad\quad
\D_{ai}(A)\ =\ \hf\Tr\, \tau_a A \tau_i A^\dagger\ .}
Here $\bra{A}$ stands for the superposition of \it explicit
\rm pointlike baryon fields (the nucleon field,
 the $\Delta$ field, and all higher spins;  a better notation
might be $\Psi^{}_A$),
any of which can be projected out using Eq.~(5.7)
below. The omitted terms, while subleading in $1/N_c,$ are needed to form
a relativistic invariant, for instance, $(\gpnn/2M_N)\partial_\mu\pi^a
\bar N\gamma^\mu\gamma^5\tau^a N$ when Eq.~\jointreq\ is projected onto
 nucleon states ($I=J=\hf$).

The most elegant result of our paper---and wholly unexpected, as this
 feature of the Skyrme model is
not present in the $U(1)$ toy model---comes from the simple requirement that
Eq.~\mombehavior\ be interpretable as a Green's function in some
quantum field theory (as we clarify at the end of this paragraph).
This requirement resolves
an operator ordering ambiguity implicit in the definition of the
pole residues ${\cal N}_1$ and ${\cal N}_2$ (the noncommuting
operators being $\bf J$ and $\D_{ai}(A)$). One finds
\eqn\Ngoeslike{{\cal N}_1\ =\ -{3i\gpnn\over2M_N}\, q^i\D_{ai}(A)
{\cal P}^{}_{\Delta J=1}\ ,\quad\quad {\cal N}_2\ =\ -{3i\gpnn\over2M_N}\,
 q^i\D_{ai}(A){\cal P}^{}_{\Delta J=0}\ .}
Here ${\cal P}^{}_{\Delta J=0}$ is the projection operator that equates
 the initial and final skyrmion spin, while
${\cal P}^{}_{\Delta J=1}$ requires that they differ by one unit, so
that any given one-pion emission or absorption
process ``sees'' only one of the two pole terms.
In this manner,  the numerators of Eq.~\mombehavior\
are brought into harmony with the LSZ interpretation of the
denominators---an interesting conspiracy between the quantum and
classical properties, respectively, of the rotationally improved
skyrmion. And the effective pion coupling to explicit pointlike
baryon fields that is equivalent to
Eq.~\Ngoeslike\ is \it precisely \rm Eq.~\jointreq. So, this paper
gives a complete solution to  the  ``Yukawa
 problem'' mentioned at the outset,
namely, showing how  Eq.~\jointreq\ emerges directly from skyrmion physics.
In retrospect, had the ${\cal P}$'s \it not \rm emerged in the numerators,
then any given one-pion process would have, in addition to a pole in the right
position, a spurious nearby isolated
pole---violating the basic precepts of the K$\ddot{\rm a}$llen-Lehmann spectral
representation of a quantum field theory,\refmark\ItZub
 and dashing any possibility of proving such an equivalence.

In related work, by
focusing on those contributions to pion-skyrmion scattering that can
be interpreted in the corresponding effective field theory as Compton
graphs, the authors of Refs.~\refs{\DiakPet\KKKO-\HSU} (a very important
precursor being Ref.~\GSII) correctly deduce pseudovector coupling with
$\gpnn\propto\gpnd\sim N_c^{3/2}.$ While we have yet to reconcile
the operator Hamiltonian formalism of Ref.~\HSU\ with our own more
pedestrian FPI approach, it appears that Refs.~\refs{\DiakPet-\HSU} are a
major step towards the complete solution, presented here, of the
``Yukawa problem.'' Another interesting idea is to extract the
effective Yukawa couplings from the skyrmion-skyrmion potential.\refmark\Oka
Additional proposed fixes to the Yukawa problem may be found in
Refs.~\refs{\VerMore\OhtaMore-\Holzwarth}.

The remainder of this
 paper is organized as follows. In Sec.~2, the nonlinear $\sigma$
model is formulated as a phase-space FPI. The baryon-number-unity
sector is then selected using a natural extension of the collective
coordinate method developed long ago by Gervais, Jevicki and
Sakita.\refmark\GJS These authors
quantized the translational mode of a one-dimensional kink, whereas
we extend the methodology to internal symmetries.
 The upshot of Sec.~2 is $SU(2)$ quantum mechanics
coupled to a quantum field theory. The latter is treated in saddle-point
approximation in Sec.~3, leading to the formal derivation of the rotationally
improved Euler-Lagrange equation \riskySUtwo. While variations on
this equation have been posited by other authors as a preferred starting
point,\refmark{\Braaten\Liuetal\Hajduk\Bander-\Rajetal}
it is reassuring to see it grounded firmly in the FPI. As Sec.~3 is
a little technical,
the reader who is already happy with Eq.~\riskySUtwo\ is encouraged
to skip directly to Secs.~4 and 5 on a first reading, as
these are the heart of our paper. In Sec.~4  we extract the large-distance
behavior of the rotationally improved skyrmion, and confirm the two
distinct poles of Eq.~\mombehavior, while in Sec.~5 we describe the
 operator ordering solution \Ngoeslike\
for ${\cal N}_1$ and ${\cal N}_2$.
In Sec.~6 the width of the $\Delta$ and of the higher-spin large-$N_c$
$I=J$ baryons are calculated. The application of rotationally improved
skyrmions to $\pi N$ scattering,\refmark{\DiakPet-\HSU,\HEHW,\MK}
 and some concluding comments, can be found in Sec.~7.

We also include four Appendices. Appendix A  revisits the $U(1)$ toy
model of Ref.~\DHM\ in a manner that more closely parallels, in a
simpler pedagogical setting, the development in
 Secs.~2 and 3. Appendix B contains a hand-waving
justification of the rotationally enhanced Euler-Lagrange
equation \riskySUtwo, and might therefore
substitute for Sec.~3 on a first pass. Appendix C discusses the effect of
the Faddeev-Popov constraints needed in the
approach of Gervais \it et al.\rm,
 while Appendix D reviews the properties of the effective
large-$N_c$ pion-baryon coupling \jointreq.

\newsec{The Skyrme Model as a Constrained Phase-Space Path Integral}

The two-flavor massive Skyrme model is defined by the Lagrangian\refmark\Skyrme
\def\calL{{\cal L}}
\def\vtau{\vec\tau}
\def\part{\partial}
\def\dag{\dagger}
\eqn\smlag{\calL\ =\ {\fpi^2\over16}\Tr\part_\mu U^\dag\part^\mu U
+{1\over32e^2}\Tr\big([U^\dag\part_\mu U,U^\dag\part_\nu U]^2\big)+
{\mpi^2\fpi^2\over8}\Tr(U-1)}
with $U$ an $SU(2)$ matrix.  The two most popular representations of the
pion field are
\eqn\oldpidef{U = \exp2i\vtau\cdot\vpi/\fpi}
or alternatively
\eqn\pidef{U=u_0+i{\vec u}\cdot\vec\tau\ ,\quad u_0^2+\vec u^2=1\ ,\quad
\vec u=2\vpi/\fpi\ .}
While these presumably define different quantum theories at ${\cal O}
(\hbar^2)$ (an underappreciated possibility\foot{This is an interesting
side-story in itself; see Ref.~\GJ\ and references therein.}$\,$),
they are equivalent for our present purposes, and we
will not need to choose between them. For
$\mpi\neq0$ the chiral symmetries $U\ \rightarrow\ AUB^\dag$
are broken explicitly to isospin, that is $B=A,$ or equivalently
$\pi^i \rightarrow \D_{ij}(A)\pi^j.$

The parameter $N_c$ enters the theory implicitly through the assignments
$\fpi^2\sim e^{-2}\sim N_c.$ Likewise, the coefficients of any desired
higher-derivative terms should also scale like $N_c,$ so that
$N_c/\hbar$ effectively sits outside the action. This observation justifies
not only the specific saddle-point calculation of Sec.~3 to follow, but also
illustrates the semiclassical picture of the large-$N_c$ world in
general.\refmark{\Witten,\GSII,\ArnMat,\ManoharRG}
 In contrast, the large-$N_c$ scaling
behavior of $\mpi$ is somewhat arbitrary. While meson masses generically
scale like $N_c^0,$ in the special case of the pion this depends
 on whether one elects
to link the chiral and large-$N_c$ limits. Since, for a reasonable
resemblance to Nature, we would like the $\Delta$ to be able to decay to $N\pi$
in our theory, and since the $N$-$\Delta$ mass difference $\sim 1/N_c,$
we will need to take the chiral limit \it at least \rm as fast as the
$1/N_c$ limit: $\mpi \sim  N_c^{-\nu}$ with $\nu\ge1.$
For technical reasons our optimal choice turns out to be
\eqn\mpiscale{\mpi\ \sim\ {1\over N_c}\ ,}
which is the convention we adopt from now on.

We remind the reader that Skyrme's choice of 4-derivative term in
\smlag\ is the unique 4-derivative construction that is at most
second order in time derivatives.\refmark\Skyrme
  This restriction is always invoked
to justify an operator quantum mechanics approach to the model.
For our present purposes, it is important
as it allows us to work in a phase-space (``Hamiltonian'') FPI
formalism, following Ref.~\GJS.
  Modulo this important restriction, we generalize
the Lagrangian \smlag\ to \it all \rm isospin-invariant models of the form
\def\V{V}
\eqn\genlag{\calL\ =\ \hf\dot\pi^ig_{ij}(\vpi)\dot\pi^j\ -\
\V(\vpi,\part_i\vpi)\ }
which admit a hedgehog soliton. The Hamiltonian is then
\def\vJ{\vec{\cal J}}
\def\calD{{\cal D}}
\def\z{\zeta}
\def\ze{\z}
\def\vz{\vec\z}
\def\calH{{\cal H}}
\def\H{\calH}
\def\ginv{g^{-1}}
\eqn\ham{\calH\ =\ \hf\z^i\ginv_{ij}\z^j+\V(\vpi,\part_i\vpi)\ ,}
where we have introduced the conjugate momenta
\eqn\momdef{\z^i\ =\ \part\calL/\part\dot\pi^i\ =\ g_{ij}\dot\pi^j\ .}

The phase-space formalism is the logically prior version of the FPI in
which one integrates over both the generalized momenta and
the generalized coordinates of the theory.\refmark\Ramond Accordingly,
the transition amplitudes between initial and final states
$\Psi_i$ and $\Psi_f,$ at times $t=-T$
and $t=+T$, in the presence of an external source $\vJ(x),$
are expressed as
\def\bigT{{\rm T}}
\def\bfx{{\bf x}}

\eqn\tfidef{\eqalign{\bigT_{fi}[\vJ\,]\ =\ \int\calD\vpi\,\calD\vec\z\,
\Psi_f^*\big[\vpi(\bfx,+T)\big]\Psi_i\big[\vpi(\bfx,-T)\big]
\exp\Big(i\int d^4x\,\z^i\dot\pi^i-\calH(\vpi,\vz)+
\vJ\cdot\vpi\,\Big)\ .}}
$\bigT_{fi}[\vJ\,]$ is the generating function  for $n$-point Green's functions
in the theory, which are extracted in the usual way
by functionally differentiating
Eq.~\tfidef\ $n$ times with respect to the external source $\vJ.$
In the current context, the advantages of the phase-space FPI are
twofold. First, it is the natural framework in which the FPI makes contact
with Hamiltonian quantum mechanics $\grave a$ $la$ Adkins, Nappi
and Witten.\refmark\ANW Second, it has the technical merit that so
long as one is careful to make a \it canonical \rm change of
variables to the collective coordinate basis,\refmark\GJS then the induced
Jacobians  cancel identically between field-space and momentum-space, as
verified below.  This is because of the  volume-preserving property of
canonical transformations. Anyone familiar with
the related topic of perturbation theory in instanton backgrounds,\refmark\LY
where the phase-space FPI is not helpful, and where
the unavoidable, uncancelled, Jacobians are best
incorporated into the Feynman rules
by means of discrete ghosts, will appreciate this simplification.

We assume  that the static Euler-Lagrange equation
\def\rhat{\hat {\bf r}}
\def\dx{d^3{\bf x}\,}
\eqn\statEul{0\ =\ {\delta \M[\vpi]\over\delta\pi^a}\ ,\quad
\M[\vpi]=\int\dx\V(\vpi,\partial_i\vpi)}
admits a  hedgehog skyrmion solution
$U_\cl = \exp\big(iF(r)\rhat\cdot\vtau\big)$.
Equivalently,
\eqn\picldef{\pi^i_\cl(\bfx)\ =\ {\fpi\over2} F(r)\rhat^i\qquad\hbox{or}\qquad
\pi^i_\cl(\bfx)\ =\ {\fpi\over2}\sin F(r)\rhat^i }
depending on which of the two parametrizations, \oldpidef\ or \pidef,
 is chosen. Isospin then generates an $SU(2)$ family of static solutions
$\D_{ij}(A)\pi^j_\cl$. Far away from the center of the skyrmion,
for any well-behaved model, $\vpi_\cl$ must be annihilated by the
static Klein-Gordon operator $\nabla^2-\mpi^2$, so that
\eqn\Fasym{F(r)\ \approx\ \sin F(r)\
\matrix{&{\scriptscriptstyle r \rightarrow\infty}\cr&\longrightarrow
\cr&{}}\
B\cdot\left({\mpi\over r}+{1\over r^2}\right)e^{-\mpi r}\ .}
For any particular choice of skyrmion Lagrangian, the numerical value
of the constant $B$ is gotten by solving the nonlinear equation for $F(r).$
Fortunately, up to chiral corrections,\foot{For the
particular choice of Lagrangian \smlag, these corrections are
${\cal O}(\mpi/e\fpi)$, which $\sim1/N_c$ by Eq.~\mpiscale.}
there is a model-independent interpretation of $B$ in
terms of the pion-nucleon coupling constant,\refmark\ANW
\eqn\Bdef{B\ =\ {3 \gpnn\over 4\pi\fpi M_N}\ ,}
that we will exploit later on.

{}From now on, we restrict the FPI \tfidef\ to configurations that live in
the baryon-number-unity sector of the theory.
In order to model physical processes involving both baryons and
mesons, we must allow for fluctuations  away
from the $SU(2)$ family of skyrmions, although still within this
topological sector.  But unless care
is taken, the resulting perturbation theory in the fluctuating
field will be plagued by infrared singularities, due to the skyrmion's
zero modes.  Specifically, the small-fluctuations operator cannot
be inverted, so the propagator is not well defined.
The cure is well known:\refmark\GJS
 one orthogonalizes the fluctuating
fields to these zero modes by means of Faddeev-Popov constraints.
To minimize clutter, in this paper we will ignore the three
translational and focus exclusively on the three rotational zero
modes $h^{\sst(k)}_m=\e_{mkl}\pi_\cl^l,$ with $k=1,2$ or 3.
 For each value of this index, the orthogonalization condition reads
\def\calO{{\cal O}}
\def\vf{\vec f}
\def\vh{\vec h}
\eqn\Ogauge{0\ =\ \calO^{\sst(k)}_\pi[A;\vpi]\ =\
\int \dx h^{\sst(k)}_m
\cdot\Big[\D_{nm}(A)\pi^n(\bfx,t)-\pi^m_\cl(\bfx)\,\Big]\ .}
A word on notation:
we will frequently abbreviate the quantity in brackets as $\delta\pi^m,$
which is the fluctuating field in the body-fixed frame of the rotating
skyrmion. Likewise, we will
 denote the body-fixed total field as $\pi_\tot^m$, defined via
\eqn\pitotdef{\pi^n\ =\ \D_{nm}(A)\cdot\big(\pi^m_\cl+\delta\pi^m\big)
\ =\ \D_{nm}(A)\cdot\pi^m_\tot\ .}
The additional incorporation of the translational modes, while
straightforward in principle, serves ultimately just to Lorentz
contract the skyrmion,\refmark\GJS which does not affect
the decay widths to leading order in $1/N_c.$ Nevertheless, for a more
accurate numerical comparison to experiment, and because it is obvious
how to do so, we will reinsert skyrmion recoil ``by hand,'' in the form
of a Lorentz-dilated skyrmion mass, in Sec.~6 below.

Formally, the three constraints \Ogauge\ are implemented by
inserting the Faddeev-Popov factor of unity into the FPI:
\def\Jpi{J_\pi}
\def\Jzeta{J_\zeta}
\eqn\FadPop{1\ =\ \int\calD A(t)\,
\det\Jpi^{ij}\cdot\prod_{k=1,2,3}\delta\big(
\calO^{\sst(k)}_\pi[A;\vpi]\big)\ .}
The definition of the Jacobian matrix $\Jpi$ depends on how
one specifies the three coordinates $a_1,$ $a_2,$ $a_3$ needed
to parametrize $SU(2)$. We will postpone making this choice explicit
for as long as possible, in which case we have, quite generally,
\eqn\Jpidef{ \Jpi^{ij}\ =\ {d\calO^{\sst(j)}_\pi\over da^i}\ =\
A^i_{nm}\int \dx h_m^{\sst(j)}\pi^n_\tot\ ,}
where
\eqn\Atensor{A^i_{nm}\  =\ {d\D_{lm}(A)\over da^i}\D_{ln}(A)\
=\ i\eps_{mnl}\Tr\big(\tau_lA^\dagger{dA\over da^i}\,\big)\ .}

Observe that \it nothing \rm in the above expressions requires that the
three constraint functions $\vh^{\sst(m)}$ be, as we originally took
them to be,  the skyrmion's rotational zero modes. They merely need to have
nonzero overlap with the zero modes, for the purpose of removing the
infrared singularities from the perturbative expansion. With this caveat,
{}from now on we will  think of the $\vh^{\sst(m)}$
 as \it arbitrary \rm functions, and will verify explicitly
that our final physical result---which cannot depend on
the division between $\vpi_\cl$ and $\delta\vpi$, as stressed in
Sec.~1---is indeed independent of the $\vh^{\sst(m)}$.
 To emphasize further
this ``gauge freedom'' we choose three  other
constraint functions $\vf^{\,\sst(m)}$ for the momentum sector, subject
only to the technical requirement that the overlap matrix
\eqn\Lambdamat{\Lambda_{ij}\ =\ \int \dx \vf^{\,\sst(i)}\cdot
\vh^{\sst(j)}}
be invertible. We denote by $P^n$ the momenta conjugate to the $a^n.$
The three momentum constraints are then
\def\vP{{\vec P}}
\eqn\momcon{0 \ =\ \calO^{\sst(k)}_\z[A,\vP;\vpi,\vz\,]\ =\
\Lambda^{-1}_{kl}\int \dx f_i^{\sst(l)}\cdot\big[
\D_{mi}(A)\z^m-\z_\cl^i\,\big]\ ,}
where the numerical prefactor $\Lambda^{-1}$ is inserted
 for later convenience, and the ``classical momentum'' $\vz_\cl(x;a_i,P_i)$
will be specified in a moment.
The corresponding Faddeev-Popov factor of unity reads
\eqn\FadPopmom{1\ =\ \int\calD \vP(t)\,
\det\Jzeta^{ij}\cdot\prod_{k=1,2,3}\delta\big(
\calO^{\sst(k)}_\z[A,\vP;\vpi,\vz]\big)\ ,\quad
\Jzeta^{ij} = {\partial\calO^{\sst(j)}_\z\over\partial P^i}
=-\Lambda^{-1}_{jl}\int\dx\vf^{\,\sst(l)}\cdot{\partial\vz_\cl\over\partial
P^i}\ .}
We will refer to the quantity in square brackets in Eq.~\momcon\ as the
body-fixed ``fluctuating momentum'' $\delta\z^i,$ and will likewise
define the body-fixed total momentum $\z^i_\tot$ analogously to
Eq.~\pitotdef.

The Faddeev-Popov insertions \FadPop\ and \FadPopmom\
effect a \it change of variables \rm in the phase-space path integral
\tfidef, from the original Lab-frame coordinates $\{\vpi(x),\vz(x)\}$ to the
far more useful set $\{\vec a(t),\vec P(t)\}\oplus\{\delta\vpi(x),\delta\vz(x)
\}$
in which the $SU(2)$ collective coordinates have been explicitly
broken out, and the remaining fluctuating degrees of freedom are expressed
in the rotating (body-fixed) frame.\foot{Generically one
expects additional ${\cal O}(\hbar^2)$ terms in the effective action
after this change of variables,\refmark\GJ but these should not affect our
leading-order result. Alex Kovner has suggested to us that
the preservation of chiral symmetry at the quantum level might forbid
such terms, as presumably they give a Yukawa-type falloff to $F(r)$,
violating Goldstone's theorem when $\mpi=0.$}
 While Eq.~\FadPopmom\ is an identity
for any choice of $\vz_\cl,$ it is particularly convenient
to choose $\vz_\cl$ in such a way that
 this change of variables is {\it canonical}, meaning that
the form of the Legendre term in the action is preserved:

\eqn\leg{\int\dx\vz\cdot{d\over dt}\vpi\ =\ P^n\dot a^n \ +\ \int\dx
\delta\vz\cdot{d\over dt}\delta\vpi\ .}
Paralleling the $U(1)$ derivation (A.8)-(A.11),
 one easily verifies
that so long as $\vz_\cl$ is  a linear combination of the three constraints
$\vh^{\sst(k)}$, and $\vz_\tot$ satisfies
\eqn\vecPdef{P^n\ =\ A^n_{jk}\int\dx\z^j_\tot\pi^k_\tot\ ,}
then the criterion \leg\ will be met. These two conditions are uniquely
satisfied by the choice
\eqn\zetacl{\vz_\cl\ =\ -\vh^{\sst(k)}\cdot
\big(\Jpi\big)^{-1}_{km}\cdot\Big(P^m-A^m_{cb}
\int\dx\delta\z^c\pi^b_\tot\Big)\ .}
As a bonus, the prefactor $\Lambda^{-1}$ in Eq.~\momcon\ then trivially
ensures that $\Jzeta^T=\Jpi^{-1}$ so that the two Faddeev-Popov determinants
 cancel  precisely in the phase-space FPI, as promised.

Thanks to the factorized form of
 the Legendre term \leg, the FPI \tfidef\ can be recast
as a \it quantum mechanical \rm sum over phase-space histories:
\def\eff{{\rm eff}}
\eqn\newtfi{\bigT_{fi}[\vJ\,]\ =\ \int\calD\vec a(t)\,\calD\vec P(t)\,
\Psi_f^*\big[A(+T)\big]\Psi_i\big[A(-T)\big]
\exp\Big(i\int_{-T}^T dt \, P^n\dot a^n\,\Big)
\exp iS_\eff[A,\vec P;\vJ\,]\ .}
Here we have anticipated the fact that to leading order in $1/N_c,$
the skyrmion wavefunctions  will be
functions of the collective coordinates only, with
no dependence on the fluctuating degrees of freedom.
For a given quantum mechanical path, the effective
action $S_\eff[A,\vec P;\vJ\,]$
is, in turn, expressible as a constrained FPI over the body-fixed fields,
\eqn\innerfpi{\eqalign{
&\exp iS_\eff[A,\vec P;\vJ\,] = \int\calD\big(\delta\vpi(x)\big)\,
\calD\big(\delta\vz(x)\big)\,\prod_{k=1,2,3}\delta\Big(\int\dx
\vh^{\sst(k)}\cdot\delta\vpi\Big)\delta\Big(\int\dx
\vf^{\,\sst(k)}\cdot\delta\vz\,\Big)\cr&\ \ \times\
\exp\Big(i\int d^4x\,\delta\z^i\delta\dot\pi^i
- \hf\z^i_\tot\cdot\ginv_{ij}(\vpi_\tot)\cdot\z^j_\tot -\V(\vpi_\tot)
 +{\cal J}^n\cdot\D_{nm}(A)\pi^m_\tot\,\Big)\ ,}}
and it is to the steepest-descent
evaluation of this expression that we now turn our attention.

\newsec{Saddle-Point Evaluation of the Effective Action}

Sufficient to leading-order in $1/N_c,$
our plan is to evaluate the inner FPI \innerfpi\ using
saddle-point methods, the goal being
 Eq.~\riskySUtwo.\foot{Appendix B might well substitute for
this rather technical Section on a first reading.} In order to do so,
 one exponentiates the $\delta$-functions in the usual
way, and extremizes the resulting effective Lagrangian
\eqn\LagSM{\eqalign{L_\eff\ =\  -\ \M[\vpi_\tot]\
 +\ \int\dx&\Big\{\delta\z^i\delta\dot\pi^i\
- \hf\z^i_\tot\cdot\ginv_{ij}(\vpi_\tot)\cdot\z^j_\tot \cr&
+\ \alpha^{\sst(k)}(t)\vh^{\sst(k)}\cdot\delta\vpi\ +\
\beta^{\sst(k)}(t)\vf^{\,\sst(k)}\cdot\delta\vz\ \Big\}\ .}}
The Lagrange multipliers $\alpha^{\sst(k)}$ and $\beta^{\sst(k)}$
implement the constraints \Ogauge\ and \momcon, respectively.
For simplicity, we are neglecting the back-reaction of the
external source $\vJ(x)$ on the saddle-point.
 This is acceptable, since the effect of
nonzero $\vJ$ can be reintroduced to any desired order in $\hbar/N_c$
using standard graphical methods.  For one-pion processes such
as $\Delta\rightarrow N\pi$ the simplest such graph is the one-loop
``lollipop'' (which is not forbidden by $G$-parity as the cubic
$\delta\vpi$ vertex is nonvanishing in the skyrmion background). Due to the
loop, this is a $1/N_c$ correction, and can be
ignored.\foot{A warning: this conclusion is special to
one-pion events. For 2-pion processes such as $\pi N\rightarrow\pi N,$
the back-reaction of $\vJ$ contributes at \it leading \rm order, and must
be taken into account; see Sec.~7.}

We look for stationary solutions to Eq.~\LagSM\
that are time independent in the rotating
(body-fixed) frame of the skyrmion, $\delta\dot\pi^i=0,$ so the Legendre term
$\int\delta\z^i\delta\dot\pi^i$ can be set to zero. Calculating
{}from Eqs.~\zetacl\ and \Jpidef\ that
\eqn\momvar{{\delta\z^i_\tot(x)\over\delta\,\delta\pi^b(y)}\ \equiv\
{\delta\z^i_\cl(x)\over\delta\pi_\tot^b(y)}\ =\
h_i^{\sst(k)}(x)(\Jpi)^{-1}_{km}A^m_{cb}\z^c_\tot(y)\ ,}
one writes down the opaque (but soon to be simplified!) intermediate
expression
\eqn\EulLag{\eqalign{0\ =\ {\delta L_\eff\over\delta\,\delta\pi^b(y)}\ &=\
-{\delta \M[\vpi_\tot]\over\delta\pi^b_\tot(y)} \ -\
 (\Jpi)^{-1}_{km}A^m_{cb}\z^c_\tot(y)
\int\dx h_i^{\sst(k)}\ginv_{ij}\z_\tot^j\cr&+\
\hf\int\dx\z^i_\tot\ginv_{ik}{\delta g_{kl}(x)\over\delta\pi^b_\tot
(y)}\ginv_{lj}\z^j_\tot\ +\ \alpha^{\sst(k)}(t)h_b^{\sst(k)}(y)\ .}}

Recasting Eq.~\EulLag\ in understandable form requires that we
eliminate all $\vz_\tot$ dependence  in favor of $\vpi_\tot$. To do so,
we once again stationarize $L_\eff$, this time
 with respect to the fluctuating momentum:
\eqn\momsta{0\ =\ {\delta L_\eff\over\delta\,\delta\z^j}\
=\ -\z^i_\tot\ginv_{ij}-\ (\Jpi)^{-1}_{kn}
A^n_{ji}\pi^i_\tot\int\dx\z^a_\tot\ginv_{ab}
h^{\sst(k)}_b\ +\ \beta^{\sst(k)}(t)f^{\sst(k)}_j\ .}
This equation is easily solved for $\vz_\tot,$ giving\foot{To
 obtain this result, multiply Eq.~\momsta\ through by
$h_j^{\sst(k)}$ and integrate to obtain $\beta^{\sst(k)}(t)\equiv0$,
and then, starting once again from Eq.~\momsta, multiply through by
$g_{jk}\pi^l_\tot A^m_{lk}$ and integrate, using
Eq.~\vecPdef\ to solve for $\int\z^a_\tot g^{-1}_{ab}h_b^{\sst(k)}$.
Observe that the saddle-point value of $\vz_\tot,$ in contrast
to $\vz_\cl,$ is ``gauge invariant,'' that is, independent of the
constraint functions.}
\def\Itilde{\widetilde\I}
\eqn\zsolve{\z^k_\tot\ =\ g_{kj}A^n_{ji}
\pi^i_\tot\Itilde^{-1}_{mn}P^m\ ,\quad
\Itilde_{mn}[\vpi_\tot]\ =\ \int\dx (A^m_{ij}\pi^i_\tot)\cdot
g_{jk}(\vpi_\tot)\cdot(A^n_{lk}\pi^l_\tot)\ .}
 Inserting Eq.~\zsolve\ into Eq.~\EulLag\ and neglecting the $\alpha^{\sst(k)}$
term for the moment, we derive the pleasingly compact variational equation
\eqn\penultimate{0\ =\ {\delta\over\delta\pi_\tot^b}\Big\{
\M[\vpi_\tot]\ +\ \hf P^m\Itilde^{-1}_{mn}P^n\Big\}\ .}
We can do even better, by reexpressing the second term using the skyrmion's
true moment of inertia,
\eqn\Idef{\I_{mn}[\vpi_\tot]\ =\ \int\dx(\eps_{ijm}\pi^i_\tot)\cdot
 g_{jk}(\vpi_\tot)\cdot(\eps_{lkn}\pi^l_\tot)\ ,}
which, unlike $\Itilde_{mn},$ is independent of the collective coordinates.
{}From Eq.~\Atensor, we obtain $\hf P^k\Omega^{-1}_{km}\I^{-1}_{mn}
\Omega^{-1}_{ln}P^l\, ,$ where $\Omega_{km}=
-i\Tr\big(\tau_kA^\dagger dA/ da^m\big)$. The reader is then invited
to choose his favorite parametrization of $SU(2)$ (ours can be found
in Sec.~5 and Appendix B)
 and verify that $P^k\Omega^{-1}_{km}=J^m,$ the angular momentum
of the skyrmion. Thus Eq.~\penultimate\ finally becomes,
\eqn\elegant{0\ =\ {\delta\over\delta\pi_\tot^b}\Big\{\M[\vpi_\tot]\
+\ \hf J^m\I^{-1}_{mn}J^n\Big\}\ ,}
subject still to the constraints $\int\vh^{\sst(k)}\cdot\delta\vpi=0.$

The results of this Section are captured in a nutshell by the
expression:
\def\Hrot{H_{\rm rot}}
\eqn\nutshell{\eqalign{\bigT_{fi}[\vJ\,]\ &\cong\
 \int\calD\vec a(t)\,\calD\vec P(t)\,
\Psi_f^*\big[A(+T)\big]\Psi_i\big[A(-T)\big]\cr&\times\
\exp\Big(i\int_{-T}^T dt \, P^k\dot a^k-\Hrot\Big)
\exp\Big( i\int d^4x\,{\cal J}^a(x)\cdot\D_{ai}\big(A(t)\big)
\pi^i_\tot({\bf x};{\bf J})\Big)\ ,}}
where $\bf J$ is the skyrmion's angular momentum, and $\Hrot$
is the rotationally enhanced energy  read off from Eq.~\elegant,
\eqn\Hrotdef{\Hrot \ =\ \M+\hf J^m\I^{-1}_{mn}J^n\ ,}
evaluated on the (constrained)
saddle-point solution $\vpi_\tot({\bf x};{\bf J})$
that minimizes $\Hrot,$ that is to say, on the rotationally improved
skyrmion.

\def\tilt{\tilde t}
\def\tilx{\tilde{\bf x}}
\def\bra#1{\left\langle #1\right|}
\def\ket#1{\left| #1\right\rangle}

\def\vev#1{\left\langle #1\right\rangle}
\def\hbfJ{\hat{\bf J}}
\def\hbfI{\hat{\bf I}}
$\bigT_{fi}[\vJ\,]$ is a generating functional for (Lab-frame)
Green's functions. In particular,  we can extract as a functional
 derivative the leading-order
amplitude for one-pion emission at the space-time point
 $(\tilt,\tilx)$ between skyrmion states $\Psi_i$ and $\Psi_f$
sharp in the collective coordinates:
\eqn\piexp{\bra{A(+T)}\pi^a(\tilt,\tilx)\ket{A(-T)}\ =\ -i
{\delta\over\delta{\cal J}^a(\tilde x)}\,\bigT_{fi}[\vJ\,]
\ \Bigg|_{\vJ\,\equiv\,\vec0}\ .}
Of course, the physical
`in' and `out' skyrmions we are really interested in are
not sharp in $A,$ but rather, sharp in spin-isospin quantum numbers.
In Sec.~6 we will review the simple rules\refmark\ANW for projecting
 out nucleons, $\Delta$'s, etc., from $\ket{A(\pm T)}.$
But already it is clear that Eq.~\piexp\ entails evaluating
$\D_{ai}\big(A(\tilt)\big)\pi^i_\tot({\tilx};{\bf J})$
between quantum states. We must therefore be prepared to answer two
questions. First, what does the rotationally improved skyrmion
$\vpi_{\rm tot}$
look like? And second, since $\bf J$ is eventually promoted by
the phase-space FPI to a Hamiltonian operator\refmark\ANW $\hbfJ,$ and since
$\hbfJ$ does not commute with the operator $\hat A,$
how is the ordering ambiguity in the
product $\D_{ai}(\hat A)\pi^i_\tot({\bf x};{\hbfJ})$ resolved?\foot{Henceforth
we put hats on quantities to denote quantum  operators.}
These questions are answered, respectively, in the two Sections to follow.

\newsec{Asymptotics of the Rotationally Improved Skyrmion}

It would appear that generating a picture of
$\pi^i_\tot({\bf x};{\bf J})$ is a complicated computational
problem, as the rotational kinetic term in Eq.~\elegant\
breaks the equivalence
between spatial rotations and isorotations (see Appendix C).
Thus the hedgehog symmetry
of the solution is spoiled, and a purely radial equation for the
skyrmion profile
is no longer available. But since we intend to focus on the
pole location in momentum space, we only need such a picture
at large distances, and here ``pure thought'' suffices.

 It is helpful to
recall some salient results from the $U(1)$ model. From Eqs.~(A.17)-(A.18),
plus the fact that in all reasonable massive models the metric
$g_{il}(\vphi_\tot\,)\rightarrow\delta_{il}$ exponentially fast at
large distances, we observe that the isorotational kinetic
term $P^2/2\I$
asymptotically contributes a negative mass-squared term $-P^2/\I^2$ to the
Euler-Lagrange equation:
\eqn\Uonenegmass{
{\delta\over\delta\phi^\tot_i}{P^2\over2\I}\ =\ -{P^2\over2\I^2}
{\delta\I\over\delta\phi^\tot_i}\ =\
 -{P^2\over\I^2}\eps_{ji}g_{jl}\eps_{lk}\phi^\tot_k\
\matrix{&{\scriptscriptstyle r \rightarrow\infty}\cr&\longrightarrow
\cr&{}}\
 -{P^2\over\I^2}\phi^\tot_i\ .}
The implications of this mass shift were reviewed in Sec.~1.

The situation for $SU(2)$ is more complicated, because the moment of
inertia $\I$ is now a tensor, and requires a little more care.
We note:
\def\tJ{\tilde J}
\def\btJ{\tilde{\bf J}}
\eqn\rotvar{{\delta\over\delta\pi_\tot^b} \hf J^m\I^{-1}_{mn}J^n\ =\
-\hf J^m\I^{-1}_{mk}{\delta\I_{kl}\over\delta\pi_\tot^b}
\I^{-1}_{ln}J^n
\ =\ -\tJ^k\eps_{ijk}\pi^i_\tot g_{jn}\eps_{bnl}\tJ^l\ ,}
adopting the shorthand
 $\tJ^k=J^m\cdot\I^{-1}_{mk}[\vpi_\tot].$ Far away from the center of
the skyrmion, we again have
$g_{jn}(\vpi_\tot)\rightarrow\delta_{jn}$ so that
\def\calM{{\cal M}}
\eqn\newasymp{{\delta\over\delta\pi_\tot^b} \hf J^m\I^{-1}_{mn}J^n\ =\
\calM_{bc}\pi^c\ ,\quad \calM_{bc}=-\btJ^2\delta_{bc}+\tJ^b\tJ^c\ .}
The mass matrix $\calM$ is  diagonalized by inspection: one nullvector
proportional to $\btJ$ itself, and two eigenstates with eigenvalue $-\btJ^2$
spanning the plane perpendicular to $\btJ,$ for which
$\btJ\times\rhat$ and $\btJ\times\btJ\times\rhat$ are a convenient
 basis. Accordingly, let us decompose $\vec\pi_\tot
=f_1\btJ+f_2\btJ\times\rhat+f_3\btJ\times\btJ\times\rhat,$ where the
$f_i$ are \it a priori \rm general functions of the invariants
$\btJ^2,$ $\rhat\cdot\btJ$ and $r$. As we have just noted,
these are constrained by the requirement
that at large distances, $f_1\btJ$ must be annihilated by $-\nabla^2+
\mpi^2$ (coming from the first term on the right-hand side of Eq.~\elegant),
while the $f_2$ and $f_3$ terms are annihilated by
$-\nabla^2+(\mpi^2-\btJ^2).$ A second constraint on the $f_i$ is the
requirement that $\vpi_\tot({\bf x};{\bf J})$ smoothly approach the hedgehog
 configuration $\vpi_\cl({\bf x})$ in the limit in which the classical
vector $\btJ\rightarrow\bf0$.
While a hedgehog  is not itself a mass eigenstate of $\calM$, it can
be formed from a superposition of the $f_1$ and $f_3$ terms, since
 $\rhat = \big((\btJ\cdot\rhat)\btJ-
\btJ\times\btJ\times\rhat\,\big)/\btJ^2$ for any $\btJ.$

Collecting the various thoughts contained in the previous paragraph,
and remembering Eqs.~\picldef-\Bdef, we write
down the following asymptotic expression for the rotationally improved
skyrmion:
\eqn\risky{\eqalign{\vec\pi_\tot({\bf x};{\bf J})\
\matrix{&{\scriptscriptstyle r \rightarrow\infty}\cr&\longrightarrow
\cr&{}}\
&{3\gpnn\over8\pi M_N\btJ^2}\cdot\Bigg\{
\Big({\mpi\over r}+{1\over r^2}\Big)e^{-\mpi r}(\btJ\cdot\rhat)\btJ
\cr&-\ \Big({(\mpi^2-\btJ^2)^{1/2}\over r}+{1\over r^2}\Big)
e^{-{(\mpi^2-\btJ^2)}_{}^{1/2} r}\,
\btJ\times\btJ\times\rhat\ \Bigg\}+\ {\calO}(\btJ)\ .}}
The ${\calO}(\btJ)$ term includes the entire contribution of the
$f_2$ term, as well as higher-order pieces from the $f_1$ and
$f_3$ terms.  We reiterate that despite the overall $1/\btJ^2,$ this
expression has a perfectly smooth limit, the hedgehog $\vec\pi_\cl,$
as $\btJ\rightarrow0.$

In the above discussion,
we have neglected the Faddeev-Popov field constraints \Ogauge\
that remain as  subsidiary conditions on Eq.~\elegant, and, one
might fear, modify $\vpi_\tot({\bf x},{\bf J})$ in some complicated
way. Fortunately, they merely  result in a rigid spatial
rotation of the skyrmion \risky\ through a small angle that vanishes
in the large-$N_c$ limit, and hence  they have no effect on the leading-order
widths. These statements are proved in Appendix C.

The $1/N_c$ expansion  allows a second simplification of Eq.~\risky,
namely, the approximation of $\I_{mk}[\vpi_\tot]$ by $\I_{mk}[\vpi_\cl]$
in the definition of $\btJ.$\foot{While this hedgehog
approximation is valid for $\I_{mn},$ it is invalid for the \it
variation \rm of $\I_{mn}$ at large $r$ as we have seen, and would
therefore miss entirely the interesting pole structure in momentum
space.}
 Since the moment of inertia tensor
evaluated on a hedgehog collapses to a scalar,
$\I_{mk}[\vpi_\cl]\equiv\I\cdot\delta_{mk},$ we simply set $\btJ\cong
{\bf J}/\I$ in Eq.~\risky. Furthermore $\I\propto\fpi^2
\sim N_c,$ which implies that the
${\cal O}(\btJ)$ contributions to $\vpi_\tot$ are also irrelevant.
Implementing these large-$N_c$ simplifications, and Fourier transforming the
rotationally improved skyrmion \risky\ to momentum space for later use,
one therefore has
\def\bq{{\bf q}}
\def\bJ{{\bf J}}
\eqn\riskymom{\eqalign{\vec\pi_\tot({\bq};{\bf J})\ &=\ -
{3i\gpnn\over2M_N \bJ^2}\cdot\Big\{ {(\bJ\cdot\bq)\bJ\over \bq^2+\mpi^2}
\ -\ {\bJ\times\bJ\times\bq\over
\bq^2+\mpi^2-\bJ^2/\I^2}\ \Big\}
\cr&+\ \hbox{($1/N_c$ corrections)}\ +\
\hbox{(non-pole terms)}\ .}}

Strictly speaking, Eqs.~\risky-\riskymom\ were derived assuming $m_\pi^2
-{\bf J}^2/{\cal I}^2>0,$ so that both exponentials are real and decaying.
Later, when we calculate the on-shell decay amplitudes, we will need
to extract the LSZ residue at $m_\pi^2-{\bf J}^2/{\cal I}^2=-{\bf q}^2.$
It is possible that the rotationally improved skyrmion itself has
bizarre properties  when this
difference goes negative,\refmark{\Braaten\Liuetal\Hajduk\Bander-\Rajetal}
such as a divergent mass at a subleading order in $1/N_c$; in fact,
there is probably no solution at all in this
regime to the defining equation \elegant. While interesting to contemplate,
and potentially useful to understand for other applications,
for our present purposes
 these pathologies are \it irrelevant$\,$\rm: Eq.~\riskymom\ will
be used to construct a Green's function with Feynman boundary conditions,
and like all such Green's functions, it is amenable to analytic
continuation.\foot{This is not unlike the instanton case.\refmark\Ringwald
There, too, the on-shell single-particle pole generated by the configuration
can only be reached with an analytic continuation away from the region
(Euclidean space) where the configuration itself is well defined.}
In sum, our procedure is: fix $m_\pi^2>0$ as per
Eq.~\mpiscale; then, since both $m_\pi^2$ and ${\bf J}^2/{\cal I}^2\sim
N_c^{-2},$ there is a finite $N_c$-independent radius of convergence
in ${\bf J}^2$ where the analysis leading to Eq.~\riskymom\ is
justified; and finally, analytically continue to the kinematic regime
of interest with the help of the usual Feynman prescription
$m_\pi^2\rightarrow m_\pi^2-i\epsilon$.

\newsec{Resolving the Operator Ordering Ambiguity}

In the previous Section, the pole pieces of the rotationally improved
skyrmion were calculated with $\bf J$ treated as a $c$-number. In
order to promote $\bf J$ to a $q$-number $\hbfJ$, one must settle the ordering
question implicit in the expression $\D_{ai}(\hat A)\pi^i_\tot({\bf x};
{\hbfJ})$,
where $\D$ is the rotation matrix from Eqs.~\pitotdef\ and \nutshell\
that relates the Lab-fixed and body-fixed frames.

\def\bfq{{\bf q}}
Such ordering ambiguities are not peculiar to the Skyrme model, or to our
particular choice of formalism, but on the contrary
appear to be unavoidable in soliton quantization. They can always be resolved
by appealing to physics. In the kink model,\refmark\GJS
the operator ordering is fixed in an  elegant way, by demanding that the
commutation relations obeyed by the generators of one-dimensional
Lorentz transformations be preserved at the quantum level.\refmark\Tomboulis
In our $U(1)$ toy model, we simply needed to invoke conservation of
energy.\refmark\DHM In both these models, the physically relevant
solution turned out to be
Weyl ordering, and hence, equivalent to the midpoint discretization
of the phase-space FPI.\refmark{\GJ,\Berezin} Unfortunately, the concepts
of Weyl ordering and midpoint discretization
do not readily generalize to $SU(2),$ which is a curved manifold (unlike
these one-dimensional examples). Nevertheless the ordering ambiguities
are easily resolved, as we now explain.

It is helpful to specify an explicit representation of $SU(2),$ namely
\eqn\suexp{A\ =\ a_0+i{\bf a}\cdot\vec\tau\ ,\qquad a_0^2+{\bf a}^2=1\ ,}
in which case
\eqn\donedef{\D_{ai}(A)\ \equiv\ \hf\Tr\,\tau_a^\dagger A\tau_i^{}
 A^\dagger\ =\ (a_0^2-{\bf a}^2)\delta_{ai}+2a_aa_i+2\eps_{aik}a_0a_k\ .}
In the $\ket{A}$ basis used by Adkins, Nappi and Witten,
 the skyrmion's mutually commuting
spin and isospin operators $\hbfJ$ and $\hbfI$ are represented
as derivatives:\refmark\ANW
\eqn\Jopdef{\hat J^k\ =\ {i\over2}\big(a_k{\partial\over\partial a_0}
-a_0{\partial\over\partial a_k}-\eps_{klm}a_l{\partial\over\partial a_m}
\big)\ =\ -{1\over4}\,\Tr\big(\tau_k\hat A^\dagger\partial_A^{}\,\big) }
and
\eqn\Iopdef{\hat I^k\ =\ {i\over2}\big(-a_k{\partial\over\partial a_0}
+a_0{\partial\over\partial a_k}-\eps_{klm}a_l{\partial\over\partial a_m}
\big) \ =\ {1\over4}\,\Tr\big(\hat A^\dagger\tau_k\partial_A^{}\,\big)\ ,}
where $\partial_A^{}=\partial/\partial a_0\,+\,i\tau_j\cdot
\partial/\partial a_j.$ Note that ${\hbfJ}^2\equiv{\hbfI}^2=
-\textstyle{1\over4}\partial_\mu\partial_\mu.$
The ordering issue arises because the components of $\hbfJ$ do not commute
with one another, nor with\foot{In fact, $\hbfJ,$ $\hbfI$ and
$\D_{ai}$ precisely generate the old $SU(4)$ spin-flavor
algebra.\refmark{\GSII,\DashMan}}
 $\D_{ai}(\hat A)$.
It is easily resolved by appealing to one fundamental property of the
spectrum of a relativistic quantum field theory, as follows.
Consider $\Delta\rightarrow
N\pi.$ As anticipated already in Sec.~1, the correct mass-shell pole will
be given by the second term on the right-hand side of Eq.~\riskymom.
As further discussed in Sec.~1, the spectral representation of
quantum field theory rules  out the possibility of an extra nearby
isolated pole in the Green's function in a theory of  pions alone.
Therefore,  the first term of Eq.~\riskymom, which  contains
such a nearby pole, cannot contribute to $\Delta$ decay, nor to any
off-diagonal transition $J_{\rm out}=J_{\rm in}\pm1.$

 We assert that the ordering\foot{N.B:
 There exist other orderings which, though they appear
distinct, give the same final result.}
 $\D_{ai}(\hat A)\hat J^i(\hbfJ\cdot\bfq)$ for the
numerator of the first term has precisely this
required property. This follows instantly from the identity
\eqn\doreyid{\D_{ai}(\hat A)\hat J^i\ =\ -\hat I^a\ ,}
since  by definition,  $\hbfJ$ and $\hbfI$ do not change the
spin/isospin representation of the skyrmion, but rather act like the
usual $SU(2)$ ladder operators within each representation.
Furthermore, we can construct a second operator that is likewise
proportional to $\hat I^a(\hbfJ\cdot\bfq)$ on any given representation,
but that has the advantage of not containing
 $\hbfI$ or $\hbfJ$ explicitly, namely,
${\cal P}^{}_{\Delta J=0}\,\D_{ai}(\hat A)q^i\ ,$ where
 ${\cal P}^{}_{\Delta J=0}$
is the projection operator that ensures that the skyrmion representation
is preserved. By the Wigner-Eckart theorem, these two operators must
therefore be proportional to one another:
\eqn\WigEck{\D_{ai}(\hat A)\hat J^i(\hbfJ\cdot\bfq)\ =\ c(\hbfJ^2)
{\cal P}^{}_{\Delta J=0}\,\D_{ai}(\hat A)q^i\ .}
The constant $c(\hbfJ^2)$ is fixed by letting both sides act on a skyrmion
wavefunction\refmark{\ANW,\Schulman}
\eqn\wvfcndef{\Big\langle\,A\ \Big|{{\scriptstyle I=J}
\atop i_z\,s_z}\Big\rangle\ =\ (2J+1)^{1/2}(-)^{J-s_z}
D^{\scriptscriptstyle(J)}_{-s_z,i_z}(A^\dagger)}
and using the standard formula for the tensor product of two
Wigner $D$-matrices.\foot{We normalize the volume of $SU(2)$ to unity.}
 One quickly  finds $c(\hbfJ^2)=\hbfJ^2,$ neatly canceling the factor of
$1/\hbfJ^2$ in front of Eq.~\riskymom.

\def\wpi{\omega_\pi}
By the same logic,
the second term in Eq.~\riskymom\ cannot contribute to the
 diagonal transitions $J_{\rm in}=J_{\rm out}.$ By rewriting
\eqn\rewriteJ{-{1\over\hbfJ^2}\hbfJ\times\hbfJ\times\bfq\ =\
\bfq\ -\ {1\over\hbfJ^2}\hbfJ(\hbfJ\cdot\bfq)\ ,}
we reduce this ordering problem to the one above, which implies
\eqn\secondterm{\D_{ai}(\hat A)\cdot\big(q^i-
{1\over\hbfJ^2}\hat J^i(\hbfJ\cdot\bfq)\big)\ =\
\big(1-{\cal P}^{}_{\Delta J=0}\big)\D_{ai}(\hat A)q^i\ =\
{\cal P}^{}_{\Delta J=1}\D_{ai}(\hat A)q^i\ .}
As the notation suggests, ${\cal P}^{}_{\Delta J=1}$
forces the skyrmion spin to change by one unit, just as required
for this term.

There is one further ordering issue to be resolved, this one not unique
to the Skyrme model, but present also in the $U(1)$ model,\refmark\DHM namely,
the meaning of $\hbfJ^2/\I^2$ in the \it denominator \rm of the second
term in Eq.~\riskymom. This is the quantity  interpreted by  LSZ
 as the squared pion energy $\wpi^2.$ As stated earlier, the
solution to this ordering question is dictated unambiguously by
conservation of energy, which equates $\wpi$  to the difference
of skyrmion energies:\foot{It
 is tempting to conjecture that for this case, the unique ordering
for $\hbfJ^2/\I^2$ specified by conservation of energy can also be
arrived at in a completely different way, by demanding that the chiral algebra
close at the quantum level, analogous to Ref.~\Tomboulis.
Note also, the lack of final-state skyrmion recoil in  this expression for the
energy difference is just a harmless byproduct of our decision at the
outset, valid to leading order in $1/N_c,$ to ignore the translational
modes.}
\eqn\massdiff{\eqalign{\wpi\ &=\ \Big(\M+{J_{\rm in}(J_{\rm in}+1)\over2\I}
\Big)\ -\ \Big(\M+{J_{\rm out}(J_{\rm out}+1)\over2\I}\Big)\ .}}
For the two allowed cases $J_{\rm in}=J_{\rm out}\pm1$, Eq.~\massdiff\
gives
\eqn\omegapiav{\wpi^2\ =\ {1\over2\I^2}\Big[J_{\rm in}(J_{\rm in}+1)+
J_{\rm out}(J_{\rm out}+1)\Big]\ ,}
that is to say, the average of $\hbfJ^2/\I^2$ acting on the bra and on the
ket (\it i.e.\rm, an anticommutator).

The various results of this Section are assembled as follows:
\eqn\opordsol{\eqalign{&
{\Big\{\D_{ai}(\hat A)\pi^i_\tot(\bfq;\hbfJ)\Big\}}_{\rm properly
\atop\rm ordered}\ =\ -{3\gpnn\over2M_N}\,\Big(
{i\over\bfq^2+\mpi^2}\cdot
{\cal P}^{}_{\Delta J=0} \,\D_{ai}(\hat A)q^i\cr&+\
{i\over\bfq^2+\mpi^2-\partial/\partial\beta}\,
e^{\beta\hbfJ^2/2\I^2}\,{\cal P}^{}_{\Delta J=1} \,
\D_{ai}(\hat A)q^i\,
e^{\beta\hbfJ^2/2\I^2}{\Big|}_{{}\atop\beta=0}\ \Big)
\cr&
+\ \hbox{($1/N_c$ corrections)}\ +\
\hbox{(non-pole terms)}\ .}}
The $\beta$ differentiation is just a concise bookkeeping device for the
nested anticommutators implied by Eq.~\omegapiav, and ensures
that $\partial/\partial\beta\,\rightarrow\,\wpi^2.$ The effective
pointlike field-theoretic vertex equivalent to this soliton expression is the
full-strength ${\cal O}(N_c^{1/2})$
pseudovector pion-baryon coupling \jointreq, as advertised,
resolving the ``Yukawa problem.''

\newsec{The Skyrmion Decay Amplitude}

\def\tilt{\tilde t}
\def\tilx{\tilde{\bf x}}
The numerical calculation of the decay widths of the $I=J$ baryons now
follow in short order. For thematic consistency,
rather than working from this point forward
with the effective field theory directly, we will complete
the calculation in the same way we started it: as a FPI dominated in
saddle-point approximation by the rotationally improved skyrmion.
But the real reason we stick with the skyrmion approach is that it is
\it much easier\rm. In the perturbative Fock space of the pion, the
skyrmion decays we are considering look like $0\rightarrow1$
transitions in a nontrivial background field, \it i.e.\rm, the skyrmion
itself, and hence the widths are integrals simply over one-body
phase space.
In contrast, the traditional relativistic effective Lagrangian approach to
$\Delta$ decay,\refmark{\Gas,\Peccei,\otherspin} not to mention the higher-spin
baryons,\refmark\otherspin  involves the construction of Rarita-Schwinger
 spinors, subsidiary spin projection conditions, and other complications.

Resuming our train of thought with Eq.~\piexp, and recalling the definition
\wvfcndef\ of the skyrmion wavefunctions, we write the cumbersome but
conceptually simple expression:
\eqn\resuming{\eqalign{&
{\phantom{\Big\rangle}}_{\rm out}\Big\langle
 {{\scriptstyle I'=J'}\atop i'_z\,s'_z}
\Big|\pi^a(\omega,{\bfq})\Big|
{{\scriptstyle I=J}\atop i_z\,s_z}  \Big\rangle_{\rm in}\ \ =
\cr&\int d\tilt\,e^{i\omega\tilt}
\int_{SU(2)}dA(T)\,
\Big\langle {{\scriptstyle I'=J'}\atop i'_z\,s'_z}\Big|\ A(T)\Big\rangle
\,\int_{SU(2)}dA(\tilt)\,
\int_{SU(2)}dA(-T)\,
\Big\langle A(-T)\ \Big|{{\scriptstyle I=J}\atop i_z\,s_z}\Big\rangle
\cr&\times\
\left(\int_{\tilt}^T\calD\vec a(t)\,\calD\vec P(t)\,
e^{i\int_{\tilt	}^T dt (P^k\dot a^k-\Hrot\,)}\right)
\cdot\bra{A(\tilt)\,}
{\Big\{\D_{ai}\big(\hat A(\tilt)\big)
\pi^i_\tot(\bfq;\hbfJ)\Big\}}_{\rm properly\atop\rm ordered}
\ket{A(\tilt)\,}
\cr&\times\
\left(\int_{-T}^{\tilt}\calD\vec a(t)\,\calD\vec P(t)\,
e^{i\int_{-T}^{\tilt} dt (P^k\dot a^k-\Hrot\,)}\right)\ .}}
Note that the FPI has been divided into two time intervals on either
side of the one-point insertion,
$-T<t<\tilt$ and $\tilt<t<T.$ A technical point: in the path integration over
each of these intervals, the quantum mechanical
field $A(\tilt)$ formally enters, wrongly, as a fixed boundary condition,
and so to lift this unphysical restriction
 we need an additional explicit integration over
 $A(\tilt)$, as we have indicated in Eq.~\resuming.
 The reason for splitting up the FPI
 in this way is that in each time segment the
skyrmion propagates \it freely \rm on the $SU(2)$ manifold. Therefore, one
can exploit the well-known sum-over-states expression for the
propagator for a free particle moving on the $SU(2)$ group manifold,
derived in a classic paper by Schulman:\refmark\Schulman
\def\rmMJ{{\rm M}_J}
\def\rmMJprime{{\rm M}_{J'}}

\eqn\Schulprop{\eqalign{&
\int_{t_1}^{t_2}\calD\vec a(t)\,\calD\vec P(t)\,
e^{i\int_{t_1}^{t_2} dt (P^k\dot a^k-\Hrot\,)}\ \ =\cr&
 \sum_{J=1/2,\,3/2,\cdots}\ \sum_{i_z,s_z=-J}^J
\Big\langle A(t_2)\ \Big|{{\scriptstyle I=J}\atop i_z\,s_z}\Big\rangle
e^{i(t_2-t_1)\rmMJ}
\Big\langle {{\scriptstyle I=J}\atop i_z\,s_z}\Big|\ A(t_1)\Big\rangle
\ ,}}
where
\eqn\MJdef{\rmMJ\ =\ \M\ +\ J(J+1)/2\I\ .}
The boundary conditions  are that $A(t_1)$ and $A(t_2)$ are held fixed.
The fact that only the diagonal component of $\I_{mn}$
appears follows from the use of the hedgehog wavefunctions $\wvfcndef,$
and is justified in large $N_c.$

Inserting Eq.~\Schulprop\ into Eq.~\resuming\ and performing the three
independent $SU(2)$ integrations using standard identities, one
extracts the much simpler expression for the one-point function,
 free of collective coordinates, independent of the choice of
Lagrangian \genlag\ and of constraints \Ogauge\ and \momcon,
and valid for both $\Delta J=0$ and $\Delta J=1$:
\eqn\freeatlast{\eqalign{&
{\phantom{\Big\rangle}}_{\rm out}\Big\langle
 {{\scriptstyle I'=J'}\atop i'_z\,s'_z}
\Big|\pi^a(\omega,{\bfq})\Big|
{{\scriptstyle I=J}\atop i_z\,s_z}  \Big\rangle_{\rm in}\ \ =\ \
-{3\gpnn\over2M_N}\,{iq^i\over\bfq^2+\mpi^2-\omega^2}
\,e^{iT\,\rmMJprime}e^{iT\,\rmMJ^{}}
\cr&\times\
(-)^{J+J'}
\big\langle J\,i_z\big|1\,J'\,a\,i_z' \big\rangle
\big\langle J'\,s_z'\big|1\,J\,i\,s_z \big\rangle
\cdot
2\pi\,\delta\big(\rmMJ-\rmMJprime-\omega\,\big)
\cr&
+\ \hbox{($1/N_c$ corrections)}\ +\ \hbox{(non-pole terms)}\ .}}
Next, one amputates the pion leg by multiplying by the inverse pion
propagator $-i(q^2-\mpi^2)$ and going on mass-shell,  killing
all non-pole contributions. The amputation of the nonrelativistic
`in' and `out' baryon legs simply means erasing the two
exponential factors, which represent free
skyrmion propagation. The conventional
definition of the $T$-matrix also requires
that we cross out the energy-conserving $\delta$-function, leaving
the product of Clebsch-Gordan coefficients,
\eqn\tmatdef{{3\gpnn q^i\over2M_N}(-)^{J+J'}
\big\langle J\,i_z\big|1\,J'\,a\,i_z' \big\rangle
\big\langle J'\,s_z'\big|1\,J\,i\,s_z \big\rangle\ .}

We now set $J'=J-1$, integrate the
square of this amplitude over the one-body relativistic phase-space
of the pion,
and sum over final states, to obtain the total skyrmion decay width:
\eqn\totwidth{\eqalign{
\Gamma_{\scriptscriptstyle J\rightarrow J-1}^{}\ =\
{\left({3\gpnn\over2M_N}\right)}^2\sum_{a,i,i_z',s_z'}
&\Big(\,\big\langle J\,i_z\big|1,J-1,a\,i_z' \big\rangle
\big\langle J-1,s_z'\big|1\,J\,i\,s_z \big\rangle\, \Big)^2
\cr&\times\ \int{d^3\bfq\over(2\pi)^3}\,{(q^i)^2\over2\sqrt{\bfq^2+\mpi^2}}
\cdot  2\pi\delta\big(\, \rmMJ\,-\,E_{\rm out}(\bfq)\,\big)\ .}}
Spherical symmetry permits $(q^i)^2\rightarrow\textstyle{1\over3}\bfq^2$
inside the integral, in which case the Clebsch-Gordan sum decouples, and
gives  $(2J-1)/(2J+1).$ The numerical value of the integral depends
sensitively on how the final-state energy $E_{\rm out}$ is defined.
With the naive choice
 $E_{\rm out}=\rmMJprime+\sqrt{\bfq^2+\mpi^2}$ as in Eq.~\freeatlast,
we obtain
\def\qbar{\bar q}
\eqn\recoilless{\Gamma_{\scriptscriptstyle J\rightarrow J-1}^{}\ =\
{3\gpnnsqd\,\qbar^3\over8\pi M_N^2}\cdot{2J-1\over2J+1}\ ,}
where $\qbar$ denotes the value of $|\bfq|$ which satisfies the
$\delta$-function. Alternatively, we can already anticipate an obvious
consequence of quantizing the skyrmion's translational as well as
rotational zero modes, namely the Lorentz dilation of the skyrmion
mass.\refmark\GJS This suggests the better
choice\foot{Because $N_c$ appears
implicitly in both the kinematics of the theory and in the parameters
themselves (unlike, say, $\alpha$ in QED), it is impossible in
practice to be a ``purist'' in the $1/N_c$ expansion, refusing to mix orders.
Nor is this even desirable in principle (a view shared by most workers
in the field), as it would break up Lorentz invariants. Those who
would object  nonetheless to our use of the mixed-order expression
$\sqrt{\bfq^2+\rmMJprime^2}$ would also need to explain how, from
the experimental values, one might separate the ``leading'' contribution
to (say) $\gpnn$ from the ``subleading'' pieces.
In our view, using the recoil-corrected skyrmion mass-energy
is truer to the spirit of equivalence to relativistic field theory
that is the main theme of this paper.}
$E_{\rm out}=\sqrt{\bfq^2+\rmMJprime^2}+\sqrt{\bfq^2+\mpi^2}\,$, in which
case the expression \recoilless\ is multiplied by an extra factor of
$\sqrt{\bfq^2+\rmMJprime^2}/\rmMJ\,=\,1+{\cal O}(N_c^{-2})$. For
$J=\thf$ these
recoil-corrected formulae give $\qbar=227\,$MeV in which
case\foot{As already noted, a comparable width of the $\Delta$ was
first quoted by Adkins, Nappi and Witten,\refmark\ANW  and similar expressions
reappear regularly  in the Skyrme-model literature.}
$\Gamma_{\scriptscriptstyle\Delta\rightarrow N\scriptstyle
\pi}\,=114\,$MeV,
against an experimental width of 120$\pm$5$\,$MeV, while the cruder estimate
\recoilless\ gives $\qbar=258\,$MeV and
$\Gamma_{\scriptscriptstyle\Delta\rightarrow N\scriptstyle
\pi}\,=212\,$MeV. In both
cases we have used the experimental values for all the parameters,
eliminating $\M$ and $\I$ from Eq.~\MJdef\
in favor of $M_N$ and $M_\Delta$ as per Ref.~\ANW\ (one of several reasonable
prescriptions).

 The recoil-corrected expressions also yield
$\Gamma_{{5\over2}\rightarrow{3\over2}}\,=803\,$MeV,
$\,\Gamma_{{7\over2}\rightarrow{5\over2}}\,=2643\,$MeV, and
$\Gamma_{{9\over2}\rightarrow{7\over2}}\,=6437\,$MeV, the masses of these
large-$N_c$ baryons being 1720, 2404 and 3284 MeV, respectively.
Extrapolating to large $J$,
 $\Gamma_{\scriptscriptstyle J\rightarrow J-1}^{}\,
\sim J^3$  while the masses grow only like $J^2$.
 So these higher-spin ``large-$N_c$
artifacts,'' often considered a failing of the Skyrme model
in particular, and
of  large-$N_c$ phenomenology in general, are so broad that they
effectively drop out of the
particle spectrum\foot{Alternatively, an interesting, purely \it
group-theoretic \rm means of eliminating the $I=J\ge5/2$ baryons from the
spectrum, while preserving unitarity, may be found in Ref.~\AmAlg.}
 for physical values of the parameters, and \it pose
no  problem whatsoever\rm.

\newsec{Application to $\pi N$ scattering, and some concluding thoughts}

By grounding the Skyrme model in the FPI, systematizing
the $1/N_c$ expansion, and paying careful attention
to the analytic properties of the rotationally improved skyrmion, we
have taken a significant step towards showing how the skyrmion bootstraps
itself into an effective relativistic quantum field theory
with explicit pointlike fields for the nucleon, $\Delta,$ etc.
 In particular, we have confirmed
using  soliton quantization\refmark\GJS the effective large-$N_c$
meson-baryon coupling \jointreq, originally put forward by Adkins, Nappi and
Witten with no explicit input from or connection to skyrmion physics,
as these authors acknowledge (see Sec.~5 of Ref.~\ANW).
By so doing, we have solved completely
the so-called ``Yukawa problem,'' namely the emergence of Eq.~\jointreq\
directly from skyrmion quantization. (Previous major progress towards
the solution may be found in Refs.~\DiakPet-\HSU.) Our approach has
the advantages of focusing on the Yukawa coupling directly (rather than
extracting it as the ``square root'' of a $\pi N$
scattering amplitude or
$NN$ potential) and being manifestly ``gauge'' ($i.e.,$ constraint)
independent. That the model-independent width of the $\Delta$ works
out well, and the problematic higher-spin baryons are too broad to
be seen, adds credibility to the large-$N_c$ program. Extensions of
our methods to the case of three flavors, and to the study of
pion photoproduction from a skyrmion, are currently in progress.

While we have focused narrowly on single-pion poles, there is obviously more
to be learned from the analytic properties of the rotationally improved
 skyrmion. For example,
look again at the asymptotic behavior \Fasym\ of the skyrmion
profile $F(r),$ obtained simply by linearizing the defining equation for $F.$
The next-leading terms, which we dropped, are ${\cal O}(F^3),$ and can
be treated in Born approximation. Their iterated contribution to the skyrmion
falls off as $e^{-\mu r},$ where the K$\ddot{\rm a}$llen-Lehmann
spectral parameter $\mu$ assumes
a continuum of values $\ge\,3\mpi.$ In the language of the corresponding
field theory, these are precisely the 2-loop vertex corrections to the bare
coupling \jointreq\ that are responsible for the 3-pion cut in the
Green's function ${\cal G}(q_\pi)$
(the 2-pion cut being forbidden by $G$ parity).
Interestingly,
the two-loop ${\cal O}(F^3)$ level is also where the two alternative
definitions of the pion field, Eqs.~\oldpidef\ and \pidef, diverge
{}from one another---and might actually define distinct quantum
theories.\refmark\GJ

Beyond diagrammatics,
a profound consequence of analyticity is crossing symmetry. We cannot
even speculate how crossing and large-$N_c$ are reconciled, since
the kinematic regimes can be so far removed from one another; for instance,
 $\Delta\rightarrow
N\pi$ with $q_\pi^2\sim N_c^{-2}$ versus the virtual transition
$\pi\rightarrow\bar N\Delta$ with $q_\pi^2\sim N_c^{2}$, the latter
being a so-called ``forbidden process'' (see Sec.~8.3 of Ref.~\Witten)
 suppressed beyond any
finite order in $1/N_c$.  Whether an obstruction to such an ``ambitious''
analytic continuation actually exists, or whether on the contrary it
is completely legitimate, is an open research topic of some importance,
and for which skyrmions might prove useful.

On a more down-to-earth level, the pion-baryon vertex is easily assembled
into more complicated diagrams, most notably $\pi N\rightarrow\pi N,$
which one might write somewhat schematically as,
\eqn\schema{\bra{\Psi_f}\D_{ai}\pi^i_\tot\big(x;\hbfJ\big)\,
\D_{bj}\pi^j_\tot\big(y;\hbfJ\big)\ket{\Psi_i}\ .}
The source of this contribution is obvious: it comes from hitting
the FPI \nutshell\ with
${\delta\over\delta{\cal J}^a(x)}\cdot{\delta\over\delta{\cal J}^b(y)}$
and pulling down two disconnected copies of the rotationally improved skyrmion.
This contribution to $\pi N$ scattering has been studied in
Refs.~\DiakPet-\HSU.
The graphs in the corresponding effective field theory are the
``Compton diagrams'' where one pion is absorbed and another emitted
directly from the baryon line.

However, there is another contribution to pion-skyrmion scattering
that has also been studied extensively in the
literature,\refmark{\KKKO,\HEHW,\MK}
which one might abbreviate as $\vev{\delta\pi^a(x)\,\delta\pi^b(y)}.$
This is the 2-point function for the fluctuating field $\delta\vpi$,
propagating in the classical background of the skyrmion, and
it contributes at the same order, $N_c^0,$ as Eq.~\schema. Yet we argued in
Sec.~1 that it is dangerous, and contrary to the semiclassical
nature of large $N_c,$  to split up the total pion field in
this way, into classical and fluctuating pieces. Is there a way of
generating this important contribution directly from the rotationally
improved skyrmion, at \it zeroth \rm order in the semiclassical
expansion?

The answer, naturally, is yes. The propagator contribution arises
automatically when
the first functional derivative pulls down the skyrmion, and
the second acts on the \it very same \rm skyrmion:
\eqn\secondcont{{\delta\over\delta{\cal J}^b(y)}
\D_{ai}\,\pi^i_\tot\big(x;\hbfJ;\big[\vJ\,\big]\,\big)
\Bigg|_{\vJ\,\equiv\,\vec0} \ .}
As the notation suggests, the rotationally improved skyrmion
is itself a functional of the
external source $\vJ(y).$ In the case of skyrmion decay, we were
cavalier in Eq.~\LagSM\ about the back-reaction of $\vJ$ on the skyrmion,
 arguing that it is a one-loop, hence $1/N_c,$ correction. But for
 $\pi N$ scattering this back-reaction is critical.
Indeed, Taylor-expanding $\D_{ai}\pi^i_\tot\big(x;\hbfJ;[\vJ\,]\,\big)$ about
$\vJ=\bf0$ we find a linear term in $\vJ,$ which is precisely the
convolution $\int dx'\,\vev{\delta\pi^a(x)\,\delta\pi^b(x')}
{\cal J}^b(x')$ of the propagator in the skyrmion background with
the source itself. This is, of course, a tree diagram, hence
leading-order, and cannot be ignored.
The associated diagrams in the effective field theory are the
``exchange-type
graphs'' in which the pion and baryon lines exchange an arbitrary number
of quanta. Thus, both contributions, \schema\ and \secondcont, can
be viewed in an elegant, unified, semiclassical
 way---in terms of the rotationally improved skyrmion.

\vskip .2in
We thank Alex Kovner and Tanmoy Bhattacharya for interesting
discussions, and Michael Peskin for commenting on the draft.
 Part of this work was completed at the Aspen Center for Physics.
Identical width calculations for the higher spin/isospin baryons have
recently been published in Ref.~\DiakNew.
We dedicate this paper to the memory of Kurt Cobain.

\appendix{A}{U(1) Redux}

In this Appendix we review some of the results of the $U(1)$ model
discussed in Ref.~\DHM, generalized to include a field-dependent
metric $g_{ij}(\vphi)$
in the kinetic energy term, so that the analogy to the Skyrme model is
closer. The incorporation of the constraints is also more general
than in Ref.~\DHM. This Appendix should be read in tandem with
Secs.~2 and 3 which it parallels closely, in a setting in which
the algebra is more transparent.

Our starting point is the 1+1 dimensional Lagrangian
\def\L{{\cal L}}
\eqn\uone{\L\ =\ \hf\dot\phi_ig_{ij}(\vphi)\dot\phi_j\ -\  V(\vphi,
\part_x\vphi)\ .}
Here $\vphi$ is a real scalar doublet, and $\L$ is presumed invariant
under the $U(1)$ transformation
\eqn\uonetrans{\vphi\ \rightarrow\ M(\theta)\cdot\vphi\ ,\quad
M(\theta)=\pmatrix{&\phantom{-}\cos\theta&\sin\theta\cr
&-\sin\theta&\cos\theta\cr}\ .}
We assume that the Euler-Lagrange equation admits
a static soliton solution $\vphi^\cl$ and hence a family of $U(1)$
solutions swept out by $M(\theta)\cdot\vphi^\cl.$\foot{There needs
to be at least one additional singlet field in order for this to be
possible,\refmark\RajWein
 but this technical point is irrelevant to our general
treatment.} As in Ref.~\GJS, we constrain the fluctuations away from these
solutions by imposing the condition
\def\O{{\cal O}}
\eqn\OgaugeUone{0\ =\ \O_\phi[\theta;\vphi]\ =\
\int dx\,h_k\cdot\big(M_{lk}\phi_l-\phi^\cl_k\big)\ .}
The expression in parentheses is the body-fixed fluctuating field
$\delta\vphi$. The constraint function $h_k$ need not be equal to the
soliton's $U(1)$ zero mode $\eps_{kj}\phi^\cl_j$; so long as they have
nonzero overlap, the constraint \OgaugeUone\ will have the desired
effect of removing the infrared singularities from the perturbative
expansion.

Formally, the constraint \OgaugeUone\ is implemented
by inserting in the path integral the Faddeev-Popov factor of unity
\def\Jtheta{J_\theta}
\eqn\FadPopUone{1\ =\ \int\calD \theta(t)\,
\Jtheta\cdot
\delta\big(\O_\phi[\theta;\vphi\,]\big)\ ,
\qquad\Jtheta\ =\ {\part\O_\phi\over\part\theta}\ =
-\int dx\,h_j\eps_{jk}\phi^\tot_k\ .}
Here the body-fixed field $\vphi^\tot$ is defined by
\eqn\phitotdef{\phi^l\ =\ M_{ln}\cdot\big(\phi_n^\cl+\delta\phi_n)\ =\
M_{ln}\cdot\phi^\tot_n\ ,}
and we have used the fact that $M_{ln}\cdot dM_{lk}/d\theta = \eps_{nk}$.

\def\H{{\cal H}}
Since we intend to use a phase-space FPI, with path integration
over both the canonical fields and their conjugate momenta, we
introduce the canonical momentum $\ze_i = g_{ij}\dot\phi_j$
in terms of which $\H = \hf\ze_i\ginv_{ij}\ze_j + V.$
We  also introduce a quantum mechanical momentum $P$ conjugate to
the $U(1)$ collective coordinate $\theta.$ The phase-space
approach\refmark\GJS requires that the momenta be constrained in analogy to
Eq.~\OgaugeUone, thus:
\eqn\momconUone{0\ =\ \O_\ze[\theta,P;\vphi,\vz\,]\ =\
\Lambda^{-1}\int dx\,f_j\cdot \big(M_{lj}\ze_l-\ze^\cl_j\big)\ .}
Here $\Lambda$ is a normalization constant that we will pick
conveniently later, and the new constraint $\vf$ can be chosen
independently of $\vh$ (so long as they have nonzero overlap).
The ``classical momentum'' $\vz^\cl$
is a configuration that we select to ensure that the constraints
\OgaugeUone\ and \momconUone\ define a canonical transformation
of the path integral variables, from the old variables $\{\vphi,\vz\,\}$
to the new variables $\{\theta,P\}\oplus\{\delta\vphi,\delta\vz\,\}$
in which the $U(1)$ collective coordinates have been explicitly
separated out. Here the body-fixed ``fluctuating momentum''
$\delta\vz$ as well as the body-fixed ``total momentum'' $\vz^\tot$
are defined as
\eqn\ztotdef{\z^l\ =\ M_{ln}\cdot\big(\z_n^\cl+\delta\z_n)\ =\
M_{ln}\cdot\z^\tot_n}
in analogy with Eq.~\phitotdef. A necessary and sufficient
condition for such a canonical transformation is that the form of
the Legendre term $\int\z_k\dot\phi_k$ be preserved, thus:
\eqn\legterm{\int dx\,\z_k\dot\phi_k\ =\ P\dot\theta\ +\
\int dx\,\delta\z_k\delta\dot\phi_k\ .}
It is an easy matter to see how the condition \legterm\ fixes $\vz^\cl$
for us, giving
\eqn\zcldef{\vz^\cl\ =\ -\vh\cdot\Jtheta^{-1}\cdot
\Big(P-\int dx\,\delta\z_k\eps_{kl}\phi^\tot_l\,\Big)\ .}
To verify this claim, expand the left-hand side of \legterm\ as follows:
\eqn\legexpand{\eqalign{\int dx\,\z_k\dot\phi_k\ &=\ \int dx\,
M_{kj}(\z^\cl_j+\delta\z_j){d\over dt}\big(M_{ki}(\phi^\cl_i+\delta\phi_i)\big)
\cr&=\ \dot\theta\int dx\,\z^\tot_k\eps_{kl}\phi^\tot_l\ +\
\int dx\,\delta\z_k\delta\dot\phi_k\ +\
\Big[{d\over dt}\int dx\,\z^\cl_k\delta\phi_k\ -\ \int dx\,\dot\z^\cl_k
\delta\phi_k\,\Big] \ .}}
Comparing the right-hand sides of Eqs.~\legterm\ and
\legexpand, we see that we must have
\eqn\Pdef{P\ =\ \int dx\,\z^\tot_k\eps_{kl}\phi^\tot_l}
while in addition the expression in square brackets in \legexpand\ must
vanish. Thanks to the constraint \OgaugeUone,
this latter condition is automatically satisfied if we pick $\vz^\cl$
proportional to $\vh,$ whence Eq.~\zcldef\ follows immediately from
the additional requirement \Pdef.

The same choice of $\vz^\cl$ that makes the Legendre term \legterm\
work out elegantly has a second nice property, as follows.
We insert into the path integral the Faddeev-Popov factor of unity
for the momentum sector:
\def\Jp{J_P^{}}
\eqn\FadPopmomUone{1\ =\ \int\calD P(t)\,
\Jp\cdot\delta\big(\O_\z[\theta,P;\vphi,\vz\,]
\big)\ ,\quad\Jp\ =\ {\partial\O_\z\over\partial P}\ =\
\Lambda^{-1}\Jtheta^{-1}\int dx\,\vh\cdot\vf\ .}
Therefore, if the normalization constant $\Lambda$ is equated to
$\int\vh\cdot\vf,$ the two Faddeev-Popov Jacobians  cancel
identically: $\Jtheta\Jp = 1.$ This reflects the volume-preserving
character of canonical transformations.

We are now ready to discuss the saddle-point evaluation of the
action. The Lagrangian reads
\eqn\LagUone{\eqalign{L_\eff\ &=\ \int dx\,\Big\{\z_k\dot\phi_k\ -\ \H\,\Big\}
\ +\ \hbox{(exponentiated constraints)}
\cr&=\  P\dot\theta\ +\ \int dx\,\Big\{
\delta\z_k\delta\dot\phi_k \ -\
\hf\ze^\tot_i\ginv_{ij}(\vphi^\tot)\ze^\tot_j\ -\ V(\vphi^\tot,
\part_x\vphi^\tot)
\cr&+\ \alpha(t)\vh\cdot\delta\vphi\ +\
\beta(t)\vf\cdot\delta\vz\ \Big\}\ .}}
Here $\alpha$ and $\beta$ are Lagrange multipliers implementing
the constraints \OgaugeUone\ and \momconUone, respectively.
We look for stationary solutions that are time independent in the
rotating frame of the soliton, $\delta\dot\phi_k=0,$ so that the
Legendre term $\int\delta\z_k\delta\dot\phi_k$ can be set to zero.
Calculating from Eqs.~\zcldef\ and \FadPopUone\ that
\eqn\momvarUone{{\delta\z^\tot_j(x)\over\delta\,\delta\phi_i(y)}
\ \equiv\ {\delta\z^\cl_j(x)\over\delta\phi^\tot_i(y)}\ =\
h_j(x)\Jtheta^{-1}\eps_{li}\z^\tot_l(y)\ ,}
one has
\eqn\EulLagUone{\eqalign{0\ =\ {\delta L_\eff\over\delta\,\delta\phi_i(y)}
\ &=\ -\Jtheta^{-1}\z^\tot_l(y)\eps_{li}\int dx\, h_jg^{-1}_{jn}
\z^\tot_n
\cr&+\ \int dx\,\Big\{\hf\z_n^\tot g^{-1}_{nk}{\delta g_{kl}(x)\over
\delta\phi^\tot_i(y)}g^{-1}_{lj}\z_j^\tot\ -\ {\delta V(x)\over
\delta\phi^\tot_i(y)}\Big\}\ + \ \alpha(t)h_i(y)\ .}}
We would like to eliminate the $\vz^\tot$ dependence of this expression,
recasting it purely in terms of $\vphi^\tot.$ To do so, we stationarize
$L_\eff$ with respect to the fluctuating momentum:
\eqn\momUone{0\ =\ {\delta L_\eff\over\delta\,\delta\z_i}
\ =\ -\z^\tot_jg^{-1}_{ji}\ -\ \Jtheta^{-1}\eps_{ij}\phi^\tot_j\int dx\,
h_ag^{-1}_{ab}\z^\tot_b\ +\ \beta(t)f_i\ .}
This equation is easily manipulated to give\foot{To
 obtain this result, multiply Eq.~\momUone\ through by $h_i$
and integrate to find $\beta(t)\equiv0,$ and then, returning to
 Eq.~\momUone, multiply through by $g_{ij}\eps_{jk}\phi^\tot_k$ and
integrate, using the identity \Pdef\ to solve for
$\int h_a\ginv_{ab}\z^\tot_b$.}
\eqn\zsolveUone{\z^\tot_k\ =\ g_{kj}\eps_{jl}\phi^\tot_l\I^{-1}P\ ,\quad
\I[\vphi^\tot]\ =\ \int dx\,(\eps^{}_{ij}\phi^\tot_j)\cdot g_{il}(\vphi^\tot)
\cdot(\eps^{}_{lk}\phi^\tot_k)\ .}
Substituting Eq.~\zsolveUone\ into \EulLagUone\ yields, finally,
the elegant expression
\eqn\elegantUone{0\ =\ {\delta\over\delta\phi^\tot_i}
\Big\{\M\ +\ {P^2\over2\I[\vphi^\tot]}\,\Big\}\ ,\quad\quad
\M\ =\ \int dx\,V\ ,}
subject still to the field constraint $\int \vh\cdot\delta\vphi=0$,
the (subleading) effect of which is discussed in Appendix C.

\appendix{B}{A hand-waving justification of the rotationally enhanced
Euler-Lagrange equation}

The purpose of this Appendix is to give an heuristic justification of
our use of Eq.~\riskySUtwo\ as an improved starting point. The reader
seeking a more compelling derivation should work through Sec.~3.

We start from the generalized Skyrme Lagrangian \genlag, and make the
ansatz
\eqn\rotansatz{\pi^i(x)\ \longrightarrow\ \D_{ik}\big(A(t)\big)
\pi^k_\tot(\bfx)}
where $\vpi_\tot$ is presumed to be time-independent. Inserting this
ansatz into the first term of Eq.~\genlag\ gives $8j^a\cdot\I_{ab}
[\vpi_\tot]\cdot j^b$, where
\eqn\littlejdef{j^a\ =\ -{i\over4}\,\Tr\,\tau^a A^\dagger\dot A}
is the $c$-number analog of the skyrmion's spin operator \Jopdef,
and $\I_{ab}$ is the moment of inertia tensor \Idef. In deriving this
result we have exploited the fact that the metric transforms as a
2-index tensor,
\eqn\twoindex{g_{ij}\big(\D\cdot\vpi_\tot\big)\ =\ g_{ab}(\vpi_\tot)
\D_{ia}\D_{jb}\ .}
Inserting the ansatz \rotansatz\ into the second term of \genlag\
just gives $V(\vpi_\tot)$ by isospin invariance. The sum of the two terms
implies an action functional that can be inserted into an ordinary
(\it not \rm phase-space) FPI. A convenient choice of coordinates
is the $S^3$-symmetric set $a_\mu$ given by Eq.~\suexp, in terms of which
the $SU(2)$-invariant path integration measure is proportional to the
product over time slices of ordinary integrals:
\eqn\prodtime{\prod_{\rm time\atop\rm slices}\int d^4a\,
\delta(a_\mu a_\mu-1)\,\exp i\int dt\,\big(8j^a\I_{ab}j^b-\M\,\big)\ .}
When $\I_{ab}$ is diagonal the first term in the exponent collapses
to $\I\,\Tr \dot A^\dagger\dot A$ which we recognize as the free $SU(2)$
Lagrangian.\refmark\Schulman

We will now show that our phase-space FPI construction,
Eqs.~\nutshell-\Hrotdef, is formally equivalent to Eq.~\prodtime.
Following Ref.~\ANW, we introduce four momenta $p_\mu$ conjugate
to the $a_\mu,$ so that $p_\mu\leftrightarrow-i\partial/\partial a_\mu,$
subject to the constraint $p_\mu a_\mu=0.$ The $SU(2)$-invariant
momentum integration is then proportional to a product over time
slices of  $\int d^4p\,\delta(a_\mu p_\mu)$. The Legendre term
can be rewritten as $\dot a_\mu p_\mu=4{\bf J}\cdot{\bf j}$
where as in Eq.~\Jopdef,
\eqn\newJdef{J^a\ =\ -{i\over4}\,\Tr\,\tau^aA^\dagger P\ ,\quad\quad
P\ =\ p_0+i\vec p\cdot\vec\tau\ .}
The phase-space FPI is then proportional to
\eqn\newprodtime{\prod_{\rm time\atop\rm slices}\int d^4a\,
\delta(a_\mu a_\mu-1)\,\int d^4p\,\delta(a_\mu p_\mu)
\exp i\int dt\,\big(4{\bf J}\cdot{\bf j}-\M-\hf J^a\I^{-1}_{ab}J^b\,\big)\ .}
Since $ d^4p\,\delta(a_\mu p_\mu)\propto d^3{\bf J}$ we
can perform the Gaussian $\bf J$ integrals and be left with Eq.~\prodtime\
precisely. This completes the heuristic justification of our phase-space
starting point.

\appendix{C}{Effect of the field constraints on the rotationally
improved skyrmion}

In Section 4, we examined the rotationally improved skyrmion while ignoring
the effect of the field constraints \Ogauge\ that remain as subsidiary
conditions on Eq.~\elegant. We justified our cavalier approach with
two claims, first, that the constraints can be implemented trivially
by rigidly rotating the configuration \risky\ in real space, and second,
that the angle of this rotation is vanishingly small
in the large-$N_c$ limit, so that the leading-order LSZ residues are
unaffected. Let us prove these two statements.

It is easiest to start with the one-dimensional $U(1)$ model
as reviewed in Appendix A.\refmark\DHM
Let $\vphi^\tot(x;P)$ be a static solution to
Eq.~\elegantUone; the configuration we have been calling $\vphi^\cl(x)$,
analogous to the undistorted hedgehog in the Skyrme model,
is then $\vphi^\tot(x;0).$ The relevant
observation is that for any $P,$ there is a $U(1)$ manifold of degenerate
static solutions, $M(\theta)\cdot\vphi^\tot(x;P).$ Since in this model
there is only one field constraint, Eq.~\OgaugeUone, generically there
will be one point, or at most a discrete set of points, on this $U(1)$
manifold that satisfy the constraint. We now show, self-consistently,
that we arrive at one such point by picking a particular
relative angle $\theta(P)$ between $\vphi^\tot(x;P)$ and $\vphi^\cl(x),$
and that $\theta(P)$ is in fact \it small\rm.
(The term ``relative angle'' presupposes that for $\theta=0,$
$\vphi^\tot(x;P)$ and $\vphi^\cl(x)$ are equivalently oriented in the
internal space, for instance to point in the $\left({1\atop0}\right)$
direction for $x\rightarrow\infty.$) The effect of a small isorotation
can be Taylor expanded:
\eqn\Uonetaylor{\phi_k^\tot(x;P)\
\matrix{&{\scriptstyle \rm isorotation}\cr&\longrightarrow\cr&{}}\
\phi_k^\tot(x;P)\ +\ \theta(P)\epsilon_{kl}\phi_l^\tot(x;P)\
+\ {\cal O}\big(\theta(P)^2\big)\ .}
The constraint becomes
\eqn\becomes{0\ =\ \int dx\,h_k\cdot\Big[\,
\phi_k^\tot(x;P)\ +\ \theta(P)\epsilon_{kl}\phi_l^\tot(x;P)\
+\ {\cal O}\big(\theta(P)^2\big)\ -\ \phi_k^\cl(x)\,\Big]}
so that
\eqn\thetapprox{\theta(P)\ \cong\ -{\int dx\, \vec h\cdot\big[\,
\vphi^\tot(x;P)-\vphi^\cl(x)\,\big]\over\int dx\,  h^{}_k\epsilon^{}_{kl}
\phi_l^\tot(x;P)}\ .}
To the extent that the rotational term $P^2/2\I$ in Eq.~\elegantUone\
is a perturbative correction to
$\M$ (as it is in the Skyrme model, where it is down by
$N_c^2$), this ratio is obviously small, and our claim is established
in a self-consistent manner.

The Skyrme model is  just a little more complicated, because of the
possibility of both spatial and isorotations in three dimensions;
respectively, $\vpi(\bfx)\,\rightarrow\,\vpi(\D\cdot\bfx)$
versus $\vpi(\bfx)\,\rightarrow\,\D\cdot\vpi(\bfx)$. The effect
of the perturbation $\hf J^m\,\I^{-1}_{mn}[\vpi_\tot]\,J^n$
is to break the $SU(2)$ degeneracy of the rotationally
improved skyrmion in $iso$space down to $U(1)$, namely isorotations in
the plane perpendicular to $\bf J.$ However, there is still a full
$SU(2)$ complement of degenerate configurations obtained by \it spatial \rm
rotations (a $U(1)$ subgroup of which is redundant with
the remaining isorotations). Thus for any $\bf J$
there is a three-parameter manifold of degenerate rotationally improved
skyrmions, and since there are now three constraints, we once again
expect a single solution point, or at most a discrete number of solutions.
So in either model the constraints, being in 1-to-1 correspondence
with the zero modes, have done their job, and eliminated
the flat directions from the FPI. The resulting perturbation
theory about the rotationally improved skyrmion will be free of IR
singularities to all orders in $1/N_c$,
 and no additional collective coordinates,
nor extra isorotational kinetic terms in the action, can be justified.
We leave it to the reader to write down the $SU(2)$ analogs of
Eqs.~\becomes-\thetapprox, which now involve matrix inverses, and
are not particularly illuminating.

\appendix{D}{Pion pseudovector coupling to the $I=J$ baryons}

As stated in Sec.~1, the pion-baryon vertex \jointreq\ is uniquely
specified by the twin requirements of the chiral and large-$N_c$
limits. The former implies the derivative coupling (``Adler's rule'').
The latter augments the usual such coupling to the nucleon,
namely
\eqn\usualstuff{(\gpnn/2M_N)\partial_\mu\pi^a\bar N\gamma^\mu\gamma^5\tau^a N\
,}
with coupling to the entire tower of $I=J$ baryons, in such a way that
the following three requirements are satisfied:

1. The coupling is invariant under isospin and angular momentum.
 The field $|A\rangle$ transforms as\foot{We remind the reader of
the compact notation of Eq.~\jointreq, whereby $\bra{A}$ is shorthand
for the superposition of \it explicit \rm pointlike
fields for the nucleon, $\Delta,$ and so forth up the $I=J$ tower of baryons,
any one of which may be projected out using Eq.~\wvfcndef.}
\eqn\eqk{
\ket{A}\
\matrix{&{\scriptstyle \rm isospin}\cr&\longrightarrow\cr&{}}\
\ket{U_I^{}A}\quad\hbox{and}\quad
\ket{A}\
\matrix{&{\scriptstyle \rm ang. mom.}\cr&\longrightarrow\cr&{}}\
\ket{AU_J^\dagger} }
so that
\eqn\eql{\eqalign{\int_{SU(2)} dA  &\,
D^{\scriptscriptstyle(1)}_{ab}(A)\ket{A}\bra{A}\ \longrightarrow\
\int_{SU(2)}dA\,
D^{\scriptscriptstyle(1)}_{ab}(A)\ket{U_I^{}AU_J^\dagger}
\bra{U_I^{}AU_J^\dagger} \cr
 = &\
\int_{SU(2)}dA\,
D^{\scriptscriptstyle(1)}_{ab}(U_I^\dagger AU_J^{})
\ket{A}\bra{A} \cr
 = &\
D^{\scriptscriptstyle(1)}_{aa'}(U_I^\dagger)
D^{\scriptscriptstyle(1)}_{bb'}(U_J^\dagger)
\int_{SU(2)}dA\, D^{\scriptscriptstyle(1)}_{a'b'}(A)\ket{A}\bra{A}\ .}}
Here we have used the group invariance of the $SU(2)$ measure, $d(U_I^\dagger
AU_J^{})=dA,$ and the reality of the rotation matrices.  Similarly,
\eqn\eqm{
\partial_b\pi^a\ \longrightarrow\ \partial_{b''}\pi^{a''}
D^{\scriptscriptstyle(1)}_{a''a}(U_I^{})
D^{\scriptscriptstyle(1)}_{b''b}(U_J^{})\ .}
Combining these last two equations confirms the invariance.

2. Equation \jointreq\ includes the usual pion-nucleon interaction
\usualstuff.  Expanding $\ket{A}\bra{A}$ into baryon fields
with good spin and isospin using the wavefunction \wvfcndef,
and performing the resulting integral over three $D$-matrices, gives
\eqn\expandout{-{3\gpnn\over2M_N}
\sum_{a,b}\partial_b\pi^a\sum_{J,i_z,s_z}\sum_{J',i'_z,s'_z}
(-1)^{J+J'}
\langle J\,1\,i_z\,a|J'\,i_z'\rangle
\langle J'\,1\,s_z'\,b|J\,s_z\rangle
\ket{J'\atop i_z'\,s_z'}\bra{J\atop i_zs_z}\ .}
We now pick out the terms with $J=J'=1/2$ in this expression.
Isospin and angular momentum invariance can be
made more manifest by rewriting this subset of terms as
\eqn\eqp{
\big(\gpnn/2M_N\big)\sum_{a,b}\sum_{i_z,s_z}\sum_{i'_z,s'_z}
\tau^a_{i_z'\,i_z}\sigma^b_{s_z'\,s_z}  \,
\partial_b\pi^a\ket{1/2\atop i_z'\,s_z'}\bra{1/2\atop i_zs_z}}
which we recognize as the nonrelativistic (or, equivalently,
in the present context, large-$N_c$) limit of Eq.~\usualstuff.

3. Equation \jointreq\ correctly implements the ``$I_t=J_t$
rule''\refmark\IJrule
and the ``proportionality rule''\refmark{\GSII,\IJrule,\DashMan}
governing meson couplings to the higher-spin fields in the
$I=J$ tower.  A careful reading of Ref.~\IJrule\ reveals these
criteria will be automatically satisfied due to the diagonality of
Eq.~\jointreq\ in the collective coordinate $A$.
It is instructive nevertheless to see how this comes about explicitly.
The bra and ket in Eq.~\expandout\ can
be written in terms of fields with good $t$-channel (exchange-channel)
quantum numbers as follows:
\eqn\eqq{\eqalign{
\ket{J'\atop i_z'\,s_z'} \bra{J\atop i_zs_z}\ = \
\sum_{I_t,I_{tz}} \sum_{J_t,J_{tz}}
(-1)^{J+i_z} (-1)^{J'+s_z'} &
\langle I_t I_{tz} | J' J i_z',-i_z \rangle
\langle JJ's_z,-s_z' | J_t J_{tz} \rangle
\cr&\times\
\ket{{I_t\,;JJ'}\atop {I_{tz}}} \bra{{J_t\,;JJ'}\atop {J_{tz}}}\ ,}}
where the phases in the above are the usual cost of turning bras into
kets in $SU(2)$:\refmark\RebSlan
 $|jm\rangle \leftrightarrow (-1)^{j+m}\langle j,-m|$.
Plugging Eq.~\eqq\ into Eq.~\expandout\  and using Clebsch-Gordan orthogonality
gives for the pion-baryon coupling:
\eqn\eqr{-{\gpnn\over2M_N}\sum_{I_{tz},J_{tz}} \partial_{J_{tz}}
\pi^{I_{tz}}\sum_{J,J'}(-1)^{J+J'}\big[(2J+1)(2J'+1)\big]^{1/2}
\ket{I_t=1\,;JJ'\atop I_{tz}} \bra{J_t=1\,;JJ'\atop J_{tz}}  \ .}

This expression correctly embodies the two aforementioned
large-$N_c$ selection rules: the square-root proportionality
factors relating the pion's couplings to the various baryon fields
 in the $I=J$ tower illustrate the  proportionality rule, while
the fact that the exchanged angular momentum $J_t=1$ (\it i.e.\rm, $P$-wave
pion emission) is equal to the isospin $I_t=1$ of the pion
is a specific example of the more general $I_t=J_t$ rule.
This latter observation is not entirely ``content-free,'' as one might
initially suspect. True, for the special case
 $\Delta\rightarrow N\pi$, or for the specific off-shell
coupling   $N\rightarrow N\pi$,
 the fact that the pion is emitted in a $P$-wave
follows trivially from parity and angular momentum conservation. But for
the higher members of the $I=J$ tower of baryons
there is no obvious conservation law forbidding, or even suppressing,
$F$-wave hard pion emission when the off-shell virtuality of the pion
is order $q_\pi^2\sim N_c^0.$ The fact that $P$-wave emission/absorption
nevertheless continues to dominate
in this kinematic regime is a specific dynamical prediction of
large $N_c,$ already incorporated into the effective
field-theoretic coupling \jointreq, and thanks to the equivalence
exhibited in this paper, also embodied by the Skyrme model. Unfortunately,
as we have also shown that these higher-spin states do not exist as
particles, this particular piece of phenomenology is somewhat moot!

\listrefs
\bye